\documentclass[aip,jcp,
floatfix,
amsmath,amssymb,
reprint,
]{revtex4-1}

\usepackage[utf8]{inputenc} 
\usepackage[T1]{fontenc} 
\usepackage{hyperref}
\usepackage{verbatim}

\usepackage[table]{xcolor} 
\usepackage{graphicx}

\usepackage{dcolumn}%
\usepackage{bm}%

\usepackage{listings}

\usepackage{algorithm}
\usepackage{algpseudocode}

\usepackage[version=3]{mhchem}
\usepackage{units}
\usepackage{siunitx}

\usepackage{xspace}

\graphicspath{{pics/}}

\renewcommand{\vec}{\boldsymbol} 

\newcommand{\eq}{Eq.\xspace}

\newcommand{\eqs}{Eqs.\xspace}
\newcommand{\fig}{Fig.\xspace}
\newcommand{\tab}{Table\xspace}

\newcommand{\Fig}{Fig.\xspace}
\newcommand{\ie}{\mbox{i.\,e.\ }}
\newcommand{\eg}{\mbox{e.\,g.\ }}

\newcommand{\Ref}{Ref.\xspace}
\newcommand{\Refs}{Refs.\xspace}
\newcommand{\lit}[1]{\Ref\citenum{#1}\xspace}
\newcommand{\lits}[1]{\Refs\citenum{#1}\xspace}

\newcommand{\SOP}{{SoP}\xspace}

\newcommand{\DOF}{{DOF}\xspace}
\newcommand{\EOM}{{EOM}\xspace}

\newcommand{\pvb}{PvB\xspace}
\newcommand{\pW}{pW\xspace}

\newcommand{\matrgreek}[1]{\ensuremath{\pmb{#1}}} %

\newcommand{\conj}[1]{\ensuremath{{#1}^\ast}}

\newcommand{\matr}[1]{\textbf{#1}}
\newcommand{\tens}[1]{\textbf{#1}}

\newcommand{\erw}[1]{\ensuremath {\langle{#1}\rangle}}

\newcommand{\braket}[2]{\ensuremath{ \langle #1 | \, #2  \rangle }}

\newcommand{\ket}[1]{\ensuremath{  | {#1} \rangle}}
\newcommand{\bra}[1]{\ensuremath{\langle {#1} | }}
\newcommand{\matrixe}[3]{\ensuremath{ \langle{#1} | \vphantom
        {#1 #3} {#2}
| {#3} \rangle }}

\newcommand{\ii}{\ensuremath{\mathrm{i}}}

\newcommand{\dd}{\ensuremath{\mathrm{d}}}

\definecolor{ocre}{RGB}{243,102,25}
\definecolor{mygray}{RGB}{243,243,244}
\definecolor{fzjred}{RGB}{175,90,80}

\definecolor{blau}{HTML}{1F78B4}
\definecolor{gruen}{HTML}{33A02C}
\definecolor{hellblau}{HTML}{A6CEE3}
\definecolor{hellgruen}{HTML}{B2DF8A}

\definecolor{nrot}{HTML}{d7191c}
\definecolor{norange}{RGB}{253,174,97}
\definecolor{ngruen}{HTML}{abdda4}
\definecolor{nblau}{HTML}{2b83ba}

\definecolor{nrot1}{RGB}{215,48,31}
\definecolor{nrot2}{RGB}{252,141,89}
\definecolor{nrot3}{RGB}{253,204,138}
\definecolor{nrot4}{RGB}{254,240,217}

\definecolor{CBred}{RGB}{215,25,28}
\definecolor{CBorange}{RGB}{253,174,97}
\definecolor{CByellow}{RGB}{255,255,191}
\definecolor{CBgreen}{RGB}{171,211,164}
\definecolor{CBlgreen}{RGB}{166,217,106}
\definecolor{CBdgreen}{RGB}{26,150,65}
\definecolor{CBblue}{RGB}{43,131,186}
\definecolor{CBblue2}{RGB}{146,197,222}
\definecolor{CBdblue}{RGB}{5,113,176}
\definecolor{CBgray60}{RGB}{102,102,102}
\definecolor{CBgray20}{RGB}{204,204,204}

\usepackage{tensor}

\newcommand{\spf}[2]{\varphi_{#2}^{(#1)}}

\newcommand{\shf}[2]{\Psi_{#2}^{(#1)}}
\newcommand{\mfm}[2]{\erw{\hat H}^{(\kappa)}_{#1#2}}

\newcommand{\densmat}[2]{\rho^{(\kappa)}_{#1#2}}
\newcommand{\densmatReg}[2]{\tilde{\rho}^{(\kappa)}_{#1#2}}
\newcommand{\proj}{\hat P^{(\kappa)}}

\newcommand{\new}{}
\newcommand{\newP}{}

\begin{document}

%
%
%
%
%
%
%
%
%
\title{Dynamical pruning of the  multiconfiguration time-dependent Hartree method (DP-MCTDH): An efficient approach for multidimensional quantum dynamics}

\author{H.~R.~Larsson}
\email{larsson@pctc.uni-kiel.de}
\affiliation{Institut für Physikalische Chemie, Christian-Albrechts-Universität zu Kiel, 24098 Kiel, Germany}
\affiliation{Department of Chemical Physics, Weizmann Institute of Science, 76100 Rehovot, Israel}
\author{D.~J.~Tannor}
\affiliation{Department of Chemical Physics, Weizmann Institute of Science, 76100 Rehovot, Israel}
\date{\today}

\begin{abstract}
We present two strategies for combining dynamical pruning with the multiconfiguration time-dependent Hartree method (DP-MCTDH), where dynamical pruning means on-the-fly selection of relevant basis functions. 
The first strategy prunes the primitive basis that represents the single-particle functions (SPFs). This is useful for smaller systems that require many primitive basis functions per degree of freedom, as we will illustrate for \ce{NO2}. 
Furthermore, this  allows for higher-dimensional mode combination and partially lifts the sum-of-product-form requirement onto the structure of the Hamiltonian,
as we illustrate for nonadiabatic 24-dimensional pyrazine.
The second strategy prunes the set of configurations of SPF at each time step.
We show that this strategy yields significant speed-ups with factors between $5$ and $50$ in computing time, making it competitive with the multilayer MCTDH method. 
\end{abstract}

\maketitle

\newcommand{\bxhTODO}{\todo[color=green!40]}
\newcommand{\bxhTODOline}{\todo[inline,color=green!40]}
\newcommand{\hrlTODO}{\todo}
\newcommand{\hrlTODOline}{\todo[inline]}
\section{Introduction}
\label{sec:introduction}
The aim of theoretical molecular quantum dynamics is to describe chemical reactions and molecular dynamics more generally by solving the time-dependent Schrödinger equation (TDSE).\cite{tannor_book} The standard approach to solve the TDSE is to expand the wavefunction in a direct product of one-dimensional basis functions. This transforms the partial differential equation into a linear algebra problem that can be solved efficiently on a computer. This simple approach works well for lower-dimensional systems but the exponential scaling with dimension of the size of the direct-product basis leads to intractable computational effort for systems with more than five-atoms.\cite{C2H_H2_zhang_2016,H2_NH2_guo_2014}

Therefore, more sophisticated methods are required to avoid the exponential scaling.
For computing (ro-)vibrational spectra, that is, solving the time-independent molecular Schrödinger equation (TISE), several techniques have been developed to circumvent the exponential scaling, see \lit{rovib_spectra_rev_carrington_2017} for a recent review. However, circumventing the exponential scaling is easier in this case  because the wavefunctions do not change in time and their shape is often simpler.

For the TDSE, the multiconfiguration time-dependent Hartree (MCTDH) method\cite{mctdh_cederbaum_1990,mctdh_NOCl_cederbaum_1992,mctdh_rev_meyer_2000} has had tremendous success. It does not eliminate the exponential scaling but  it reduces the base. The method employs a time-dependent direct-product basis that is evolved variationally at each time-step.
These basis functions are called single-particle functions (SPFs).
The MCTDH method allowed for the simulation of 12-dimensional systems.\cite{CH4_H_manthe_2001}
It leads to significant reduction of computational effort, but only for weakly coupled systems (between weakly coupled groups of modes).
Moreover, the direct application of the MCTDH \emph{Ansatz} is still affected by exponential scaling. This was alleviated by so-called ``mode combination'', where a direct-product basis of not one- but higher-dimensional basis functions are used and variationally optimized.\cite{pyrazine_24d_cederbaum_1998,mctdh_rev_meyer_2000} 
With this, even 24-dimensional problems were tractable already almost 20 years ago.\cite{pyrazine_24d_cederbaum_1998,pyrazine_24d_cederbaum_1999}
More recently, the MCTDH \emph{Ansatz} has been used to describe higher-dimensional SPFs, leading to the multilayer MCTDH method (ML-MCTDH).\cite{ml_mctdh_thoss_2003,ml_mctdh_manthe_2008,ml_mctdh_meyer_2011,ml_mctdh_rev_wang_2015}
For model systems, the treatment of hundreds or even thousands of degrees of freedom (\DOF) is then possible.\cite{ml_mctdh_thoss_2003,ml_mctdh_meyer_2011,ml_mctdh_anthracene_c60_lan_2015,ml_mctdh_FMO_kuehn_2016}
Despite its success, the MCTDH method with mode-combination and its ML variant have their drawbacks. Finding optimal combinations of modes or ML-``trees'' is difficult.\cite{ml_mctdh_H_CH4_manthe_2012,mlmctdh_versus_mctdh_H2COO_meyer_2014,ml_spawning_gatti_2017} Furthermore, the MCTDH approach is only advantageous if the Hamiltonian is in the form of a sum of products (\SOP) of operators acting on only one degree of freedom.\cite{mctdh_rev_meyer_2000} Fitting the Hamiltonian into this form is possible but adds an additional layer of complexity and approximation:\cite{potfit_meyer_1996,multigrid_potfit_meyer_2013,mctdh_multilayer_potfit_otto_2014,pes_neural_network_sum_of_products_carrington_2006,sop_pes_neural_networks_zhang_2014,pes_sum_of_products_smolyak_carrington_2015,pes_sop_from_multimode_fit_rauhut_2016}
Fitting the Hamiltonian into \SOP form for higher-dimensional systems is still challenging.
With mode-combination, the Hamiltonian may be expressed as a sum of products of operators acting on more \DOF.\cite{zundel_mctdh_hamiltonian_meyer_2007} Correlation DVR is an alternative but it is approximate and challenging to use within the context of ML-MCTDH.\cite{cdvr_manthe_1996,ml_mctdh_manthe_2008,ml_cdvr_2009,mctdh_denmat_inv_manthe_2015}

Building on previous ideas,\cite{proDG_hartke_2006,proDG_hartke_2006,pruning_mccormack_2006,pruning_wyatt_2006,pruning_wyatt_2007,pvb_edyn_takemoto_tannor_2012,pvb_edyn_tannor_2015} we have recently proposed an alternative method to implement the TDSE for higher-dimensional systems.\cite{pW_tannor_2016} We use the standard \emph{Ansatz} of a time-independent direct-product basis but employ basis functions that lead to a sparse representation of the wavefunction. The basis functions are either coordinate-space-localized discrete-variable-representation (DVR) functions,\cite{dvr_rev_light_1992,tannor_book} phase-space-localized projected von Neumann functions, \pvb, \cite{semicl_gauss_heller_1979,pvb_tannor_2012,pvb_edyn_takemoto_tannor_2012,pvb_math_tannor_2016,pvb_rev_tannor_2017} or phase-space-localized but momentum-symmetric projected Weylets, \pW.\cite{pW_tannor_2016,gabor_basis_wilson_1987,wilson_gabor_basis_journe_1991,weylet_1_poirier_2003,weylet_2_poirier_2004,weylet_3_poirier_2004,weylets_Ne2_poirier_2006}
If the wavefunction is expanded in one of these bases, large parts of the wavefunction coefficient tensor have negligible amplitude, \ie they are sparse and the values are below a certain threshold. The corresponding terms in the wavefunction expansion can then be dropped, \ie the basis is pruned. 
The time-dependence requires an adaptive scheme where basis functions are added and removed during the propagation of the wavepacket. 
The three mentioned bases can be pruned very efficiently. DVR and \pW turned out to be useful also for higher-dimensional systems. 
We showed the efficiency of our approach for up to six-dimensional systems~\cite{pW_tannor_2016} and we believe that this pruning strategy can be an alternative to MCTDH for systems with intermediate dimensionality. Because the pruning is not based on exploiting weak correlation, it might be especially useful for highly-correlated systems. Despite the tremendous decrease of effort compared to unpruned dynamics, simple pruning of a direct-product grid will not suffice for applications with more than about nine or twelve \DOF without further development. This is due to the exponential increase of the size of the boundary of the wavefunction in higher dimension.\cite{asaf_thesis}
Note, however, that systems with more than twelve \DOF can indeed be treated within this approach if (ro-)vibrational spectra are of interest. \cite{pvb_LiCN_tannor_2014,pvb_H2O_carrington_2015,pvb_math_tannor_2016,pvb_rev_tannor_2017,symmetr_gauss_appl_CH2NH_poirier_2014,symmetrized_gaussians_acetonitrile_poirier_2015,benzene_hybrid_truncation_scheme_HO_basis_poirier_2015,symmetrized_gaussians_sop_carrington_2016}

Clearly, pruning can be combined with the MCTDH \emph{Ansatz}, because, in principle, any direct-product Hilbert space can be pruned, as long as the representation of the wavefunction is sufficiently sparse. There are two ways to use the pruning strategy within (ML-)MCTDH. Both will be considered in this contribution.
The SPFs are normally described by a so-called primitive direct-product basis.
Thus, the first way to combine pruning with the MCTDH approach is to prune this primitive basis. This allows for a straightforward combination of our pruning methodology from \lit{pW_tannor_2016} with the MCTDH method. It might be useful for SPFs that require many primitive basis functions. Moreover, pruning the SPF representation allows for higher-dimensional mode combination, making MCTDH more favorable for higher-dimensional systems.
Furthermore, higher-dimensional mode coupling relaxes the requirements regarding the form of the Hamiltonian.
Pruning is particularly useful for complicated and highly correlated dynamics. 

Note that the primitive bases DVR, \pvb and \pW,\cite{pW_tannor_2016} which we will use for pruning the SPF representation, are grid-based methods. As an alternative, one could also regard the two-layer version of G-MCTDH\cite{G-MCTDH_Burghardt_1999} as a compact representation of the SPFs.\cite{G-MCTDH_layer_Burghardt_2013} G-MCTDH, however, adds an additional layer of complexity. 

The second way is to prune the wavefunction in SPF representation, \ie the MCTDH coefficient tensor. Often, the most time-consuming parts of a MCTDH computation is the handling of the coefficient tensor, and therefore the savings can be significant. Consequently, this idea is not new and was first succesfully implemented by Worth.\cite{mctdh_selected_configurations_worth_2000} It can be regarded as an alternative to ML-MCTDH but it can also be combined with the ML-MCTDH strategy. The earlier cousins of the MCTDH methods, namely  multiconfiguration time-dependent self-consistent field \emph{Ansätze},\cite{mctdscf_miller_1987,mctdscf_kosloff_1987,mctdscf_kosloff_1990} might also be considered as pruned MCTDH methods.\cite{mctdh_selected_configurations_worth_2000}

The work of Worth was extended by Haxton and McCurdy in the context of electron dynamics.\cite{restricted_mctdh_variational_principle_mccurdy_2015} The time-dependent Restricted-Active-Space Self-Consistent-Field (TD-RASSCF)\cite{tdRASSCF_madsen_2013,tdRASSCF_extension_madsen_2014,tdRASSCF_space_partition_madsen_2017} methodology  and the occupation-restricted multiple-active-space model (TD-ORMAS)\cite{tdORMAS_ishikawa_2015} may also be viewed as pruned MCTDH methods for fermions. The former has recently been formulated for bosonic systems.\cite{td_rasscf_bosons_madsen_2017}

Variational Multiconfigurational Gaussians, vMCG,\cite{vMCG_Burghardt_2004,vMCG_rev_lasorne_2015} might be considered as a radical way to prune the SPF space. There, the SPFs are not a direct-product basis but a selection of time-dependent Gaussians.\cite{G-MCTDH_Burghardt_1999} However, this method is challenging to implement numerically.\cite{vMCG_rev_lasorne_2015}

In the context of computing vibrational levels,  \ie solving the TISE, one can use the MCTDH \emph{Ansatz} within the so-called ``improved diagonalization'' approach developed by Meyer \emph{et al.}\cite{mctdh_improved_diagonalisation_gatti_2016} 
This approach can be considered as vibrational Complete-Active-Space Self-Consistent-Field, vCASSCF\cite{vCASSCF_lievin_1995} or as a special case of vibrational Multi-Configurational Self-Consistent-Field, vMCSCF.\cite{vMCSCF_lievin_1994,vMCSCF_rauhut_2010}
Pruning the configurational space in vMCSCF has been thoroughly studied by Rauhut \emph{et al.} and Mizukami and Tew,\cite{vMCSCF_rauhut_2010,vMCSCF_pruned_rauhut_2010,vMRPT2_pruned_tew_2013,vMCSCF_pruned_state_averaged_rauhut_2015}
including a further perturbative treatment.\cite{vMRPT2_pruned_tew_2013,vMCSCF_pruned_perturbation_theory_rauhut_2014}
Recently, also Wodraszka and Carrington followed this path and presented an efficient algorithm for the required pruned tensor transformations.\cite{pruned_mctdh_carrington_2016}

In all the above methods the SPF space is not pruned dynamically and the size of the pruned space is not allowed to change. In this contribution, we introduce, for the first time, dynamic pruning as presented in \lit{pW_tannor_2016} for standard time-independent direct-product bases into MCTDH, called DP-MCTDH. This greatly increases the effectivity of pruning. 
By defining a wave amplitude threshold $\theta$, which determines the accuracy of the pruning, one avoids the need to predetermine the number of SPFs in each dimension. Essentially, this yields the MCTDH method with just one parameter. 
This might be an alternative to ML-spawning where the parameters in ML-MCTDH are determined dynamically, and again one effective parameter is used.\cite{ml_spawning_gatti_2017} \new{We compare DP-MCTDH against conventional MCTDH and ML-MCTDH using well-established benchmark systems.}

\new{After submitting this paper, Wodraszka and Carrington published another article about pruned vCASSCF (MCTDH with improved diagonalization).\cite{wodraszka_2017} 
There, they use an adaptive pruning for solving the TISE, similar in spirit of \lit{pvb_algorithms_tannor_2016}. 
They refined their algorithm from \lit{pruned_mctdh_carrington_2016} for the pruned matrix-vector product using our ideas from \lit{pW_tannor_2016} and compare their method with ML results. There are three key differences from the present work. 1) They do not prune the primitive basis, just the coefficient tensor; 2) They do not use mode combination; 3) In contrast with the present paper, they are not solving the TDSE.  Using adaptive or dynamic pruning for solving the TDSE is more difficult than using it for the TISE because the basis has to be adapted at each time-step. This makes it necessary to develop efficient algorithms for updating the pruned basis.\cite{pW_tannor_2016} Additionally, the error of the pruning depends on the wavefunction at previous times, whereas, for solving the TISE, a pruned bases can always be refined without the dependence on previous results. By analyzing reduced densities, we will show why it is important to use dynamical/adaptive instead of static pruning.}

In the following, we briefly restate the MCTDH theory in Section \ref{sec:MCTDH_theory}. Then, we describe the pruning of the SPF representation and of the coefficients in detail in Sections \ref{sec:theory_prune_SPF} and \ref{sec:theory_prune_A}, respectively. More details regarding our implementation are given in Section \ref{sec:implementation}. This is followed by examples in which the primitive basis, is pruned (Section \ref{sec:ex_NO2}), the set of configurations of SPFs is pruned (Section \ref{sec:ex_pyr}) and where both pruning strategies are combined (Section \ref{sec:ex_combined}). The reduced density in DP-MCTDH is analyzed in Section \ref{subsec:pyr_anal}.
We conclude in Section \ref{sec:conclusion}.

\section{Theory}
\subsection{General MCTDH theory}
\label{sec:MCTDH_theory}
In MCTDH,\cite{mctdh_cederbaum_1990,mctdh_NOCl_cederbaum_1992,mctdh_rev_meyer_2000} the $D$-dimensional wavefunction $\ket{\Psi(t)}$ is expanded in a direct-product of a so-called single-particle basis $\{\ket{\spf\kappa{j_k}(t)}\}_{j_k=1}^{n_\kappa}$:
\begin{align}
 \ket{\Psi(t)} &= \sum_{j_1=1}^{n_1}\sum_{j_2=1}^{n_2}\dots \sum_{j_D=1}^{n_D} A_{j_1j_2\dots j_D}(t) \bigotimes_{\kappa=1}^D \ket{\spf\kappa{j_k}(t)},\\
 &\equiv \sum_{J} A_J(t) \ket{\Phi_J(t)},
\end{align}
where we have introduced the multi-index $J = j_1j_2\dots j_D$. Both the coefficient tensor $\tens A$ and the single-particle functions (SPFs) $\ket{\spf\kappa{j_k}}$ are time-dependent. Mathematically, this corresponds to a Tucker decomposition of the full tensor of $\ket{\Psi}$ in a primitive basis representation (see below).\cite{tensor_decomp_rev_bader_2009} In the context of quantum chemistry, this form is known as Complete-Active-Space Self-Consistent-Field, CASSCF.\cite{helgaker_book} 
Throughout the text, we use atomic units unless stated otherwise. 

Inserting this \emph{Ansatz} into the time-dependent Schrödinger equation, $\ii \partial_t \ket{\Psi(t)} = \hat H \ket{\Psi(t)}$ , employing the Dirac-Frenkel-McLachlan variational principle,\cite{td_variational_principle_dirac_1930,td_variational_principle_frenkel_1934,td_variational_principle_mclachlan_1964,td_var_principle_equivalence_van-leuven_1988} $\matrixe{\delta \Psi}{\ii \partial_t - \hat H}{\Psi} = 0$,
and restricting the variations of the SPFs to satisfy 
\begin{align}
      \braket{\spf\kappa i}{\spf \kappa j} &\stackrel!= \delta_{ij}\quad \text{and} \\
     \braket{\spf\kappa i}{\partial_t\spf{\kappa}{j}} &\stackrel!= 0
\end{align}
gives\cite{mctdh_cederbaum_1990,mctdh_NOCl_cederbaum_1992,mctdh_rev_meyer_2000}
\begin{align}
\ii \partial_t A_J &=   \sum_L \matrixe{\Phi_J}{\hat H}{\Phi_L}A_L \equiv \sum_L H_{JL} A_L,\label{eq:mctdh_havec}\\
\ii \partial_t \ket{\spf\kappa j} &= (\hat 1-\proj) \sum_{k,l=1}^{n_\kappa} [\densmatReg jl]^{-1} \mfm lk \ket{\spf \kappa k}.\label{eq:mctdh_spf_eq}
\end{align}
$\proj$ projects onto the space spanned by the SPFs:
\begin{equation}
 \proj = \sum_{i,j=1}^{n_\kappa} \ket{\phi_i}[S_\phi^{-1}]_{ij}\bra{\phi_j}, 
\end{equation}
with $[S_\phi]_{ij} = \braket{\spf\kappa i}{\spf\kappa j}$. Initially, $\matr S_\phi$ is a unit matrix but the limited precision of the numerical solver of the system of ordinary differential equations (ODE) leads to spurious nonorthogonalities that need to be taken into account.\cite{mctdh_rev_meyer_2000}

Introducing the so-called single-hole functions
\begin{equation}
 \ket{\shf\kappa l}\equiv \braket{\spf\kappa l}\Psi,
\end{equation}
the so-called mean-field matrix and density matrix can be written as
\begin{align}
  \mfm j l &= \matrixe{\shf \kappa j}{\hat H}{\shf \kappa l},\label{eq:mctdh_mf_mats} \\
    \densmat j l &= \braket{\shf \kappa j}{\shf \kappa l}.\label{eq:mctdh_dens_mats}
\end{align}

The rank of the $n_\kappa\times n_\kappa$ density matrix, $\matrgreek \rho^{(\kappa)}$, might be less than $n_\kappa$ and an inversion is not possible. To solve this issue, the density matrix is regularized using\cite{mctdh_rev_meyer_2000}
\begin{equation}
 \tilde{\matrgreek \rho}^{(\kappa)} = \matrgreek \rho^{(\kappa)} + \epsilon \exp(-\matrgreek \rho^{(\kappa)} / \epsilon).\label{eq:mctdh_regularization}
\end{equation}
Depending on the parameter $\epsilon$, this regularization can decrease the efficiency of the ODE solver for too low values and can sometimes cause inaccuracies and instabilities.\cite{mctdh_denmat_inv_inacc_bonitz_2016}
Recently, new formulations that either ``hide'' the rank-deficiency\cite{mctdh_denmat_inv_lubich_2015,mctdh_denmat_inv_lubich_2017} or remove it using a perturbative treatment\cite{mctdh_denmat_inv_fischer_2014,mctdh_denmat_inv_manthe_2015} have been developed.

To solve the equations for the SPFs, \eq\eqref{eq:mctdh_spf_eq}, the SPFs are expanded in a so-called primitive basis $\{\ket{\chi^{(\kappa)}_{a}}\}_{a=1}^{N_\kappa}$:
\begin{equation}
\ket{\spf\kappa i(t)} = \sum_{a=1}^{N_\kappa} U^{(\kappa)}_{ai}(t) \ket{\chi^{(\kappa)}_{a}}. \label{eq:spf_expansion}
\end{equation}
The primitive basis is normally a time-independent discrete variable representation (DVR) basis.\cite{dvr_rev_light_1992,tannor_book,dvr_origin_heenen_1986,lagrange_mesh_rev_baye_2015} Other bases like \pvb can be used as well. In the latter case, the primitive basis becomes nonorthogonal and the overlap matrix of the primitive basis has to be included in the equations of motion (\EOM) for the matrix $\matr U^{(\kappa)}$.

The memory-requirement is proportional to $D n N + n^D$, where $n$ is the geometric mean of the set of the numbers of SPFs, $\{n_\kappa\}_{\kappa=1}^D$, and $N$ is similarly defined for the number of primitive basis functions. If we assume that the Hamiltonian, $\hat H$, can be decomposed as a sum of products of one-dimensional operators (see Section \ref{sec:tensor_trafo}), the computational effort is proportional to $e_1 D n N^2 + e_2 D^2 n^{D+1}$, where $e_1$ and $e_2$ are some coefficients.\cite{mctdh_rev_meyer_2000} Note the exponential scaling with respect to the number of SPFs.  This scaling limits the applicability of MCTDH to $D \lesssim 12$.

If so-called mode combination is employed,\cite{pyrazine_24d_cederbaum_1998,mctdh_rev_meyer_2000} some degrees of freedom are described by higher-dimensional single-particle functions. The primitive basis functions, $\{\ket{\chi^{(\kappa)}_{a}}\}_{a=1}^{N_\kappa}$, are then multidimensional.
This decreases the dimension of the coefficient tensor $\tens A$ and hence leads to a reduction in computational effort and storage requirement. However, the effort and storage needed for the description of the single-particle functions is increased.\footnote{For mode combination, the scaling in computational effort for the description of the SPFs can be implemented as $g D n^2  M^{P+1}$, where $P$ is the dimension of the SPFs and $M$ is the geometric mean of the numbers of one-dimensional primitive functions describing the SPFs. This is often more favorable than a $g D n M^{2P} = g D n N^2$ scaling.}
Nevertheless, if not too many modes are combined, the overall computational effort and storage requirement can be reduced. 
This is \new{because the} required number of multidimensional SPFs is typically less than the product of the required numbers of one-dimensional SPFs, especially if highly correlated modes are combined. 
Mode combination led to the successful simulation of 24-dimensional pyrazine,\cite{pyrazine_24d_cederbaum_1998,pyrazine_24d_cederbaum_1999} see also \new{Sections \ref{sec:ex_pyr} and \ref{sec:ex_combined}}.
In ML-MCTDH,\cite{ml_mctdh_thoss_2003,ml_mctdh_manthe_2008,ml_mctdh_meyer_2011} the multidimensional SPFs are described recursively using the MCTDH approach itself, which corresponds mathematically to a hierarchical Tucker decomposition.\cite{hierarchical_tucker_decomp_kuehn_2009,hierarchical_tucker_decomp_grasedyck_2010}
Finding the best mode combination can be a nontrivial task.\cite{ml_mctdh_H_CH4_manthe_2012,mlmctdh_versus_mctdh_H2COO_meyer_2014,ml_spawning_gatti_2017}

\subsection{Pruning within the MCTDH method}
There are several ways to apply pruning to the MCTDH method and the different strategies can be combined. One can prune the primitive basis used for expanding the SPFs and/or the set of configurations of the SPF basis, i.~e., the coefficient tensor. 
These pruning strategies are described in the following.
\subsubsection{Pruning the primitive basis}
\label{sec:theory_prune_SPF}

The most straightforward application of possibly multidimensional wavefunction pruning as introduced in \lit{pW_tannor_2016} to MCTDH consists in pruning the representation of the (multidimensional) SPFs, i.~e., to prune the expansion in \eq\eqref{eq:spf_expansion} like
\begin{equation}
\ket{\spf\kappa i(t)} = \sum_{a\in \mathcal A_i^{(\kappa)}} U^{(\kappa)}_{ai}(t) \ket{\chi^{(\kappa)}_{a}},\label{eq:spf_expansion_pruned}
\end{equation}
where $\mathcal A_i^{(\kappa)}$ is the subset of employed indices. Because the shapes of the SPFs change in time, the subsets should be time-dependent.
For mode combination, $a$ can be a multi-index and the primitive basis functions $\ket{\chi^{(\kappa)}_{a}}$ can be a Hartree-product of one-dimensional functions.
For efficient pruning, a judicious choice of primitive functions is paramount.
In \lit{pW_tannor_2016} we benchmarked coordinate-space-localized DVR functions, phase-space-localized projected von Neumann functions, \pvb, and phase-space-localized momentum-symmetrized projected Weylets, \pW.
We found that \pvb gives the most compact representation but exhibits in higher dimensions an unfavorable computational scaling with respect to the basis size. The representation in \pW is less compact but the scaling is as good as in the DVR. However, the latter was found to have a smaller prefactor in the scaling due to the diagonality of the matrix representation of the potential, even though the representation is less compact than \pW.

In many applications of MCTDH, the propagation of the SPFs is not the most time-consuming part. However, this changes if many primitive basis functions in one (combined) mode are required. This is typically the case in lower-dimensional systems with photodissociation or reaction dynamics for at least one or two coordinates. For larger systems, it may happen if many modes are combined. By pruning the SPFs, the effort required to propagate the SPFs is decreased and it becomes possible to combine more modes to propagate even higher-dimensional SPFs.

\subsubsection{Pruning the configurations of single-particle basis functions}
\label{sec:theory_prune_A}
In most applications, the propagation of the coefficient tensor, $\tens A$, and the set-up of the mean-field matrices, \eq\eqref{eq:mctdh_mf_mats}, are the most time- and memory-consuming parts.  It is therefore natural to try to prune the configurational space spanned by the SPFs. This was successfully done by Worth,\cite{mctdh_selected_configurations_worth_2000} including a careful analysis of the consequences of pruning. See Section \ref{sec:introduction} for further examples where this approach or variants of it have been used.

Until now, only static pruning conditions have been used to prune MCTDH \new{for solving the TDSE (see \lit{wodraszka_2017} for adaptively pruned  vCASSCF)}. However, the number of required configurations that contribute to the wavefunction normally changes (and often increases) during the propagation. Furthermore, it may happen that previously important configurations become negligible at later times.
Therefore, static pruning is not the best choice, although the easiest to implement. Instead, we use our dynamic pruning method described in \lit{pW_tannor_2016}; see also \lits{proDG_hartke_2008,pvb_edyn_tannor_2015,pvb_algorithms_tannor_2016}:
At each time-step, all configurations with a coefficient magnitude larger than a specified wave amplitude threshold, $\theta$, get nearest neighbors in configuration space added. All configurations that have magnitudes smaller than $\theta$ and that have no nearest neighbors with magnitudes larger than $\theta$ are discarded. 

If natural SPFs (natural orbitals), \ie SPFs or orbitals that diagonalize the density matrices, \eq\eqref{eq:mctdh_dens_mats},\cite{mctdh_natural_orbitals_jensen_1993,mctdh_natural_orbitals_manthe_1994,mctdh_rev_meyer_2000} are used, the SPFs can be ordered by their natural population or weight. This provides for a well-defined order of configuration space.
In the context of electronic structure theory, it was shown that natural orbitals give the sparsest representation.\cite{natural_orbitals_lowdin_1955}

Because our selection of significant configurations is dynamical, the overall number of required SPFs generally changes with time. Hence, there is no requirement to specify the number of SPFs, $n_\kappa$, in advance. This means that pruned MCTDH has essentially just one parameter, the wave amplitude threshold.

Haxton and McCurdy noticed that in an arbitrarily pruned MCTDHF method, the different variational principles are not identical.\cite{restricted_mctdh_variational_principle_mccurdy_2015} They tried different  approaches to find the best \EOM in a reduced configurational space. Here, we simply follow Worth\cite{mctdh_selected_configurations_worth_2000} and use the MCTDH \EOM as described in Section \ref{sec:MCTDH_theory} without further modifications. 
To the extent that our dynamic pruning scheme selects the most important configurations up to a given threshold $\theta$, the standard MCTDH equations will be sufficient up to that chosen level of accuracy.

\section{Implementation}
\label{sec:implementation}
We have implemented the MCTDH and the DP-MCTDH methods in a new computer code. The basis, the Hamiltonian and the initial state are set up using the Python programming language. This is then interfaced to a compiled code, written in the C++ programming language. 
The object-orientation of the two languages allowed for a simple implementation of the pruned and unpruned variants of our code without creating significant overhead. 
For the unpruned parts of the code, we use the Eigen linear algebra library\cite{eigen_lib}  that is interfaced against the Intel® MKL library.\cite{mkl_lib} The tensor transformations (see next Section \ref{sec:tensor_trafo}) of both the unpruned and pruned bases follow \lit{pW_tannor_2016}. Our (DP-)MCTDH implementation is as fast or slightly faster than implementations in the Fortran programming language but currently not parallelized.

\subsubsection{Pruned tensor transformation}
\label{sec:tensor_trafo}
An efficient implementation of matrix-vector products or tensor transformations $\matr H \vec A$ as found in \eqs~\eqref{eq:mctdh_havec} and \eqref{eq:mctdh_mf_mats} is of pivotal importance.
Assuming  $\hat H$ to be decomposed in a sum of direct products of one-dimensional operators (\SOP), 
\begin{equation}
 \hat H = \sum_{l=1}^g c_l \bigotimes_{\kappa=1}^D  \hat h^{(\kappa,l)},\label{eq:sop_structure}
\end{equation}
the tensor transformation scales as $g n^{D+1}$ instead of $n^{2D}$.\cite{DVR_matvec_calc_friesner_1986,DVR_matvec_calc_manthe_1990,DVR_matvec_calc_carrington_1993} By introducing one approximation, this scaling is retained in the pruned case, even if the pruning is without any structure. We will review this procedure as introduced in \lits{pW_tannor_2016,pvb_rev_tannor_2017} briefly.
Assuming a two-dimensional Hamiltonian like 
\begin{equation}
\hat H = \hat h^{(1)} \otimes \hat h^{(2)},
\end{equation}
one can write it as 
\begin{equation}
\hat H = (\hat h^{(1)} \otimes \hat 1^{(2)}) (\hat 1^{(1)} \otimes \hat h^{(2)}),
\end{equation}
or, in matrix representation,
\begin{equation}
\matr H = (\matr h^{(1)} \otimes \matr 1^{(2)}) (\matr 1^{(1)} \otimes \matr h^{(2)}) = \matr H^{(1)} \matr H^{(2)}.
\end{equation}
$\matr H$ might be dense but $\matr H^{(\kappa)}$ are permuted block-diagonal matrices of size $n^{D} \times n^{D}$. Each block has a size of $n\times n$. Thus, successive multiplication of  $\matr H^{(\kappa)}$ with $\matr A$ takes $n^{D+1}$ operations.

Pruning can be introduced by a matrix $\matr R$ that is a rectangular matrix of size $n^D \times \tilde{n}^D$ with ones on the diagonal, where $\tilde n$ is the size of the \emph{pruned} one-dimensional basis. Projection into the pruned subspace then gives 
\begin{align}
 \matr R^\dagger \matr H \matr R &= \matr R^\dagger \matr H^{(1)} \matr H^{(2)} \matr R \\
 &\approx  [\matr R^\dagger \matr H^{(1)}  \matr R] [\matr R^\dagger \matr H^{(2)} \matr R] = \widetilde{\matr H}^{(1)} \widetilde{\matr H}^{(2)}.
 \label{eq:pruned_sequantial_matvec}
\end{align}
The last approximation is used in our implementation;\cite{pW_tannor_2016,pvb_rev_tannor_2017} see also \lit{pruned_prod_basis_mapping_carrington_2009}, where this concept was first used for a pruned basis. 
This type of approximation is known as the product approximation in DVR theory.\cite{dvr_rev_carrington_2007,tannor_book,pruned_prod_basis_mapping_carrington_2009} It leads to nonhermitian matrices\cite{pvb_rev_tannor_2017,pruned_prod_basis_mapping_carrington_2009} but the error introduced by the approximation is generally lower than the error introduced by the pruning. 

This approximation reintroduces the favorable scaling of order $\tilde{n}^{D+1}$. We emphasize that no arrays of size larger than $\tilde{n}^D$ are required and that the number of basis functions in each dimension can differ.
No scaling with respect to the size of the unpruned tensor is involved. In practice, no matrix  $\widetilde{\matr H}^{(\kappa)}$ is explicitly stored. However, we need to permute the vectors between the successive application of $\widetilde{\matr H}^{(\kappa)}$. This simplifies the algorithm and makes each operation local in memory.\cite{pW_tannor_2016} \new{By allowing intermediate arrays of size larger than $\tilde{n}^D$ (and thus a slightly less favorable scaling), the product approximation can be avoided.\cite{pruned_mctdh_carrington_2016,wodraszka_2017} We found the product approximation not to be severe, see above and Section \ref{sec:impl_prune_A}.}

\new{The computation of mean-field and density matrices, \eqs \eqref{eq:mctdh_mf_mats} and \eqref{eq:mctdh_dens_mats}, can be done using the same permutation strategy. The density-like matrix $\matr D$, $D_{lm}^{(\tau)} = \braket{\Psi^{(\tau)}_l}{\bar{\Psi}^{(\tau)}_m}$, can be evaluated as $\matr A^\dagger \bar{\matr A}$, where $\matr A$ is the matricized and permuted pruned coefficient tensor of size $\left(\prod_{\kappa\neq \tau} \tilde n_\kappa\right) \times \tilde n_\tau$.
In our implementation, both the tensor and the matricized tensor are stored in memory as one-dimensional arrays \emph{of the same size}. The only difference is the dimension that is represented contiguously in memory. No product form is required.
More details are given in \lit{pW_tannor_2016}. Only a minor adjustment of the algorithm described in the Appendix of \lit{pW_tannor_2016} is needed to compute density matrices. Here, no approximation is involved.}

\subsubsection{Pruning the primitive basis}

Each SPF typically occupies a slightly different region in phase space. Therefore, the most compact representation is obtained by using different subsets $\mathcal A_i^{(\kappa)}$ for each single-particle function $i$ in mode $\kappa$. However, to ease the implementation, we use a common subset for all SPFs in one (combined) mode. It turns out that this still leads to a compact representation.

After each time-step, we prune the primitive basis again. Once primitive basis functions have been removed or added (by setting their coefficient to zero), the SPFs are no longer orthogonal and Löwdin's symmetric orthogonalization procedure is used to restore orthogonality.\cite{lowdin_orthogonalization_lowdin_1950} We have tested different orthogonalization methods and found no significant instability caused by the orthogonalization. We prefer Löwdin orthogonalization because it retains the shape of the SPFs as closely as possible.

\subsubsection{Pruning the configurations of single-particle basis functions}
\label{sec:impl_prune_A}
\paragraph{Propagation}
We closely follow Worth\cite{mctdh_selected_configurations_worth_2000} and use a constant mean-field (CMF) propagator.\cite{mctdh_cmf_meyer_1997}
The Bulirsch-Stoer solver\cite{bulirsch-stoer_book2} is used for propagating the SPFs and the short-iterative Lanczos solver\cite{sil_light_1986,mctdh_rev_meyer_2000} for propagating the coefficient tensor. Both have been taken from the Heidelberg package\cite{mctdh_package} and interfaced against our C++ code.

After each time step of the propagator, we transform to natural SPFs (natural orbitals).\cite{mctdh_natural_orbitals_jensen_1993,mctdh_natural_orbitals_manthe_1994,mctdh_rev_meyer_2000}
Changing the propagation equations to fulfill the natural orbital requirement turned out to be numerically unstable and required more time steps,\cite{mctdh_selected_configurations_worth_2000} even if a variable mean-field propagator\cite{mctdh_rev_meyer_2000} is used. 

Once we transform the orbitals into the natural orbital representation, the orbitals are ordered by their occupation number or natural population. Hence, our dynamical pruning scheme might be too general because nearest neighbors are added in all directions in configuration space, regardless of the natural population. An improvement might be considered in the future.

\paragraph{Pruning}
The SPFs are pruned before each transformation into natural orbitals.
Our dynamical pruning does not assume any structure; therefore, our pruned tensor transformation cannot use the algorithms described in \lits{mctdh_selected_configurations_worth_2000,pruned_mctdh_carrington_2016} without jeopardizing the favorable scaling of the transformation. We used our implementation from \lit{pW_tannor_2016}, which is based on permutations; see Section \ref{sec:tensor_trafo}.  The cost of the additional permutation operations was negligible in our examples in \lit{pW_tannor_2016}. It is not negligible in our examples considered in Sections \ref{sec:ex_pyr} and \ref{sec:ex_combined}, because few SPFs are required in each dimension and there are many unit operators in the \SOP form of the considered Hamiltonians. In our original\new{, cyclic permutation} scheme, we need to permute the basis in order to apply the unit operation \new{in dimension $\tau$} and afterwards, permute it again to apply the next operator \new{in dimension $\tau-1$. This requires a storage of $D$ vectors of permutation indices. To avoid unnecessary permutations, we introduce noncyclic permutations and thereby reduce the computational cost for applying unit operators. This is implemented in practice by saving} all possible permutations between each dimension, storing $(D-1)(D-2)/2$ additional vectors of the size of the \emph{pruned} coefficient tensor. Wodraszka and Carrington also needed to adapt their algorithm to handle Hamiltonians with many unit operators.\cite{pruned_mctdh_carrington_2016} 

We note that our pruned tensor transformation leads to nonhermitian mean-field matrices; see Section \ref{sec:tensor_trafo}.
Nevertheless, the numerical propagation turned out to be stable while an explicit symmetrization of the matrices actually decreased the stability. If a symmetrized pruned product of type $[\widetilde{\matr H}^{(1)} \widetilde{\matr H}^{(2)} + \widetilde{\matr H}^{(2)} \widetilde{\matr H}^{(1)}] / 2$ (compare with \eq\eqref{eq:pruned_sequantial_matvec}) is used,\cite{pruned_prod_basis_mapping_carrington_2009} hermiticity is restored but the error of the product approximation in the matrix-vector product remains and is only slightly decreased. Hence, we did not use this symmetrized product.
Hermiticity is attained by increasing the size of the reduced SPF subspace.

\paragraph{Newly added SPFs}
Dynamical pruning requires the removing and adding of SPFs. Removing SPFs is not difficult but the initial representation of newly added SPFs can be problematic because they do not contribute to the wavefunction, at the given timestep. Mendive-Tapia \emph{et al.} used a random representation and orthogonalized the SPFs afterwards.\cite{ml_spawning_gatti_2017} We prefer to use the first Krylov vectors obtained by multiplying the uncorrelated part of the Hamiltonian with the SPFs in that dimension. We have also tried other Krylov spaces like that generated by the position operator but found no significant advantage.
The optimal way would be to use the scheme by  Lee and Fischer\cite{mctdh_denmat_inv_fischer_2014} and Manthe,\cite{mctdh_denmat_inv_manthe_2015} but it is computationally expensive as the scaling is $\mathcal O[(N^{(\kappa)})^2]$ for each degree of freedom $\kappa$.

\new{All steps involved in the pruning procedure are summarized as a pseudo-code in the Appendix.}

\section{Examples}
\subsection{Pruning the primitive basis: \ce{NO2}}
\label{sec:ex_NO2}

To test the pruning of the primitive basis, we use the example of \ce{NO2} dynamics on the $B_2$ surface\cite{NO2_mctdh_cederbaum_1992} and closely follow \lit{pW_tannor_2016}. This three-dimensional example has been claimed to exhibit ergodic dynamics. Many (20 in each coordinate) SPFs are needed to correctly describe the autocorrelation function. Further, many primitive functions are needed: We use 250 Fourier Grid Hamiltonian functions\cite{fgh_marston_balint-kurti_1989,tannor_book} in the radial coordinates and 100 Gauss-Legendre-DVR functions in the angular coordinate; see \lit{pW_tannor_2016} for further details of our set-up.
For this system, more than 60\% of the computing time is spent on solving the SPF \EOM.

We compare the pruning of both coordinate-space-localized DVR and phase-space-localized \pvb. The latter is formed by a similarity transformation of the DVR basis. Without pruning, it leads to exactly the same results (within machine accuracy).\cite{pvb_tannor_2012,pvb_math_tannor_2016,pW_tannor_2016} We have previously shown that \pvb is not optimal for higher-dimensional dynamics due to a less favorable scaling.\cite{pW_tannor_2016} Here, no mode combination is used and the SPFs are one-dimensional. Therefore, the scaling in higher dimensions does not matter. 

\begin{figure}
 \includegraphics{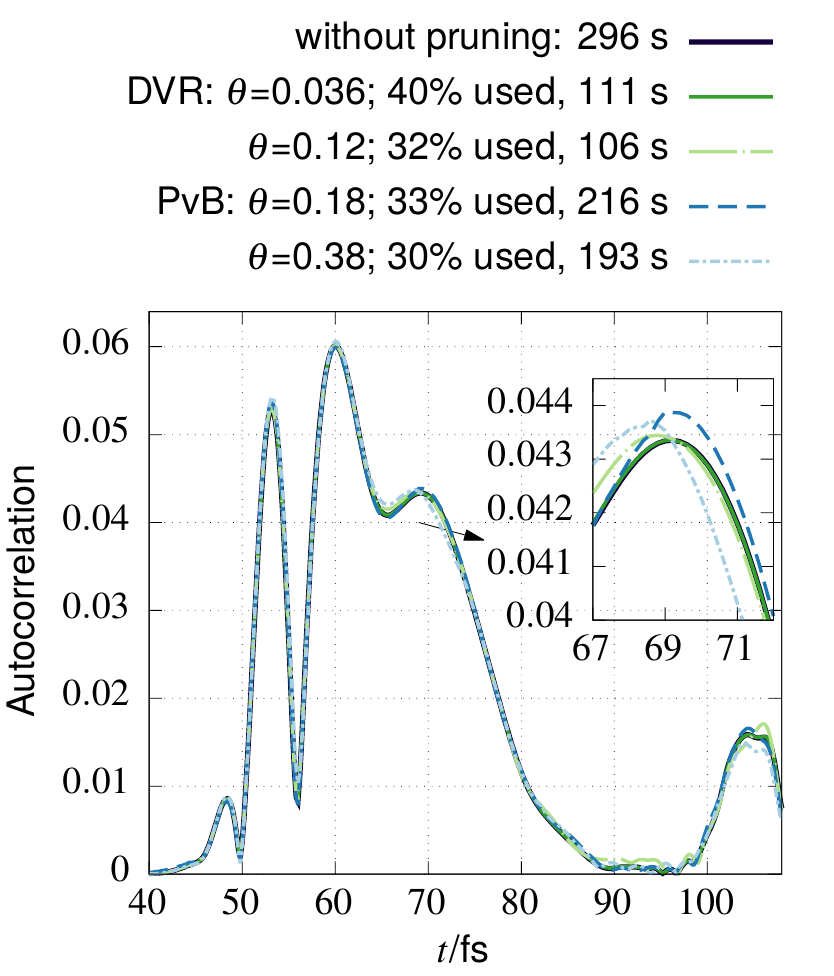}
 \caption{Autocorrelation function of \ce{NO2} dynamics without pruning (black) compared to pruned DVR (green lines) and pruned \pvb (blue lines) dynamics. The percentages denote the average fraction of the total basis size used for all coordinates. 100\% corresponds to a basis size of ($256+256+100$). The times denote the overall runtime and $\theta$ denotes the wave amplitude threshold.}
 \label{fig:NO2_acorr}
\end{figure}

\Fig \ref{fig:NO2_acorr} shows the autocorrelation function ($\braket{\Psi(0)}{\Psi(t)} = \braket{\conj{\Psi(t/2)}}{\Psi(t/2)}$) obtained from unpruned dynamics (using the DVR basis), and pruned dynamics using either DVR or \pvb. The accuracy of the CMF propagator was set to $10^{-6}$. The computation was done on a single core of an Intel(R) Core(TM) i7-6700 processor. We used the same wave amplitude threshold, $\theta$, in all coordinates. %
Using different thresholds for the angular and for the radial coordinates would maybe give an improvement. Note that the values of $\theta$ for a requested accuracy level depend on the dimensionality and the choice of basis. Better definitions of $\theta$ might be considered in the future. \new{In all of our pruned benchmark calculations, we simply scanned various values of $\theta$ on a logarithmic scale to find appropriate values.}
Throughout, we use a regularization parameter of $\epsilon=10^{-8}$ (see \eq\eqref{eq:mctdh_regularization}).

The unpruned dynamics (black line) took 296 seconds runtime. 
If 40\% of the overall basis is used on average (green line), pruned DVR can accurately reproduce the unpruned dynamics and 62\% of the overall runtime can be saved ($\unit[111]s$ runtime). \pvb is less efficient. If 33\% of the overall basis is used (dashed blue line), the autocorrelation function is less accurately reproduced and only 27\% of runtime is saved ($\unit[216]s$ runtime). This is a result of the nondiagonality of the potential matrices in the \pvb representation, see \lit{pW_tannor_2016}.

\subsection{Pruning the configurations of single-particle basis functions: Pyrazine}
\label{sec:ex_pyr}
Here, we consider the quadratic vibronic-coupling model of 24-dimensional (plus an electronic degree of freedom, \DOF) pyrazine.\cite{pyrazine_24d_cederbaum_1998,pyrazine_24d_cederbaum_1999}
To simplify the implementation, we use the so-called single-set mode\cite{mctdh_rev_meyer_2000} where the same SPF basis is used for both electronic states. The SPFs are expanded in Gauss-Hermite-DVRs.\cite{dvr_origin_heenen_1986,lagrange_mesh_rev_baye_2015,tannor_book} 
Including the electronic \DOF, the coefficient tensor, $\tens A$, is nine-dimensional.

The basis parameters and mode combinations are shown in \tab\ref{tab:pyr24_basis}. 
There, we show two variants for the number of SPFs, A and B. Variant A corresponds to the numbers used in \lits{pyrazine_24d_cederbaum_1999,mctdh_selected_configurations_worth_2000} whereas variant B corresponds to the ``MCTDH-3'' variant from \lit{ml_mctdh_meyer_2011}. Many more SPFs are used in variant B. We closely follow \lit{pyrazine_24d_cederbaum_1999} and \new{first} use variant A \new{that also allows for a comparison with \lit{mctdh_selected_configurations_worth_2000} (Section \ref{subsec:pyr_varA}).} The spectrum computed from the dynamics with variant A is already converged to a reasonable level but the autocorrelation function is not. Therefore, we also show results from variant B and later compare also with ML-MCTDH results \new{(Section \ref{subsec:pyr_varB}). Comparing variants A and B provides insights into the scalability of DP-MCTDH. If the reader is more interested in the general performance and comparison to ML-MCTDH, the studies on variant A (Section \ref{subsec:pyr_varA}) may be skipped.}
Finally, we present unrestricted simulations (no limit regarding the number of SPFs) in Subsection \ref{subsec:pyr_noRestriction}.

\begin{table}
\caption{Mode combination and basis parameters of the (24+1)-dimensional pyrazine example. Two variants for the number of SPFs are shown, see the text for details.}
\label{tab:pyr24_basis}
\begin{ruledtabular}
 \begin{tabular}{llll}
 \DOF & Combined normal modes & Number of SPFs & Primitive \\ 
                   &    & variants \{A, B\}     & basis size \\ \hline
0 & electronic & \{2,2\} & 2\\
1 & $[\nu_{10a}, \nu_{6a}]$ & \{14,21\} & $[40,32]$\\
2&  $[\nu_1, \nu_{9a}, \nu_{8a}]$ & \{8,12\} & $[20,12,8]$\\
3&  $[\nu_2, \nu_{6b}, \nu_{8b}]$ &\{6,7\} & $[4,8,24]$\\
4& $[\nu_4,\nu_5,\nu_3]$ & \{6,8\} & $[24,8,8]$\\
5& $[\nu_{16a}, \nu_{12},\nu_{13}]$ & \{5,7\} & $[24,20,4]$\\
6& $[\nu_{19b}, \nu_{18b}]$ & \{7,12\} & $[72,80]$\\
7& $[\nu_{18a}, \nu_{14}, \nu_{19a}, \nu_{17a}]$ & \{5,7\} & $[6, 20, 6,6]$\\
8& $[\nu_{20b}, \nu_{16b}, \nu_{11}, \nu_{7b}]$ & \{4,5\} & $[6,32,6,4]$
 \end{tabular}
 \end{ruledtabular}
\end{table}

\subsubsection{Variant A (fewer SPFs)}
\label{subsec:pyr_varA}

For variant A, almost 95\% of runtime is spent on routines where $\tens A$ is involved. Hence, pruning only $\tens A$, that is, the SPF space, makes sense. Because the unpruned propagation is not fully converged  with respect to the number of SPFs, we have limited the maximal number of SPFs in the pruned dynamics to the number of SPFs in the unpruned dynamics.

We stress that variant A is not converged with respect to the number of SPFs, $n_\kappa$. Compared to variant B, the autocorrelation function differs. The restricted values for $n_\kappa$ in variant A lead to a strong dependence of the dynamics on the choice of the initially unoccupied SPFs. Depending on their shape, the SPFs and thus the wavefunction ``drift'' into different spaces. %
In particular, values of the autocorrelation function at the maxima are sensitive to this choice, whereas the positions of the maxima are not.
Even minor perturbation of the unoccupied SPFs have a strong influence on the autocorrelation function at later times. The sensitivity of the values can be considered as an indicator of the accuracy. 
The better converged variant B is much less sensitive to the choice of the initial SPFs. There, a perturbation yields almost no difference in the autocorrelation function for the first \unit[80]{fs}. For larger times, the difference is less than for variant A.

\begin{figure*}
 \includegraphics{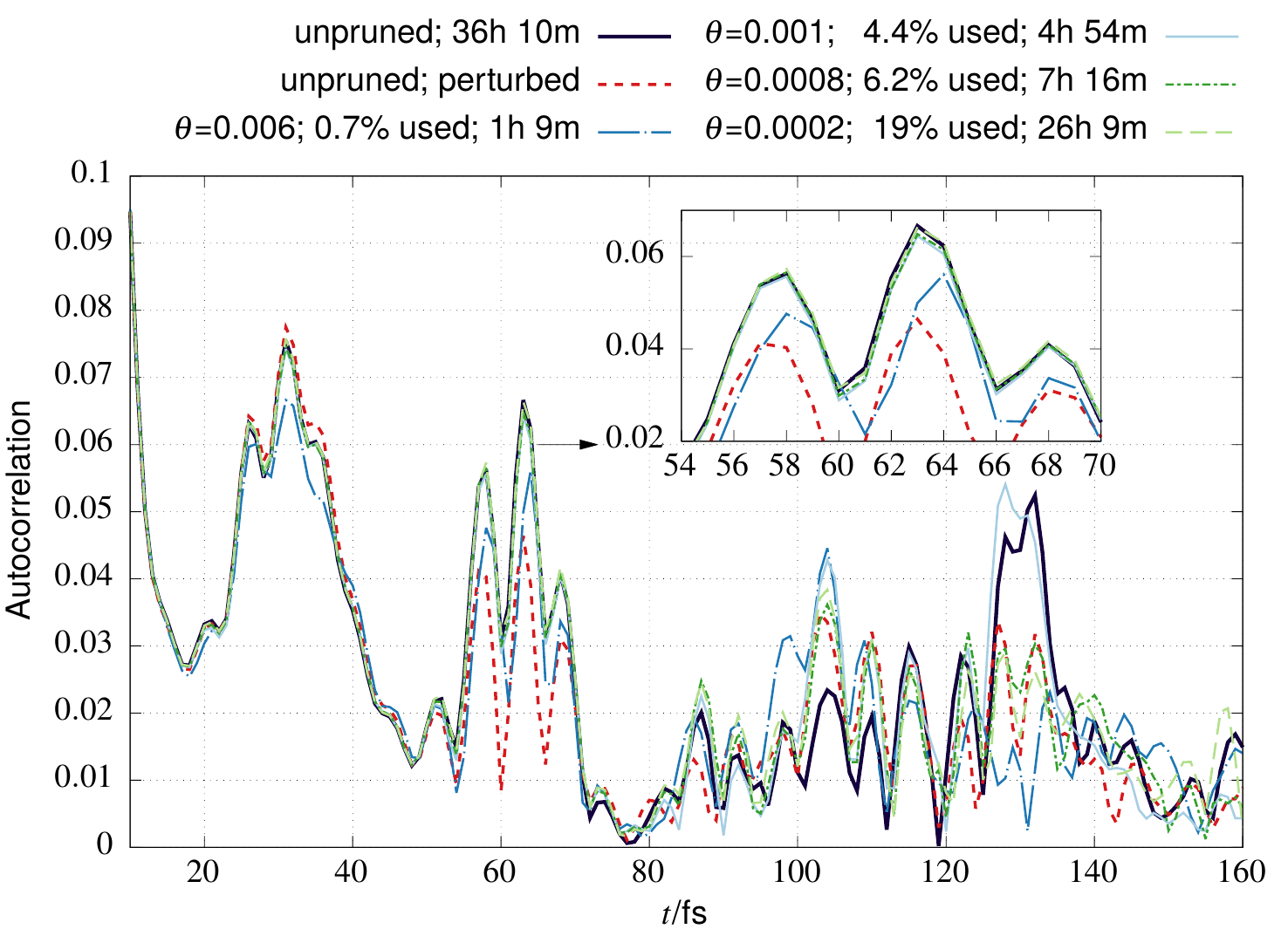}
 \caption{Autocorrelation function of unpruned pyrazine dynamics compared to dynamics with a pruned coefficient tensor. %
 \newP{The (maximal) number of SPFs corresponds to variant A in \tab \ref{tab:pyr24_basis}.}
 The percentages denote the average size of the pruned configuration space used. 100\% corresponds to a size of $5.6\cdot 10^{6}$. The times denote the overall runtime and $\theta$ denote the wave amplitude threshold. The dashed red curve shows unpruned dynamics where the initial and unoccupied SPFs are randomly perturbed and illustrates the general accuracy of the unpruned dynamics.
 }
 \label{fig:pyr24_A_acorr}
\end{figure*}

Following the implementation in the Heidelberg MCTDH package,\cite{mctdh_package} we prepare the initially unoccupied SPFs by generating an orthonormal Krylov space of the operator $\hat x$: $\ket{\spf\kappa j}= \hat x^{j-1} \ket{\spf\kappa1}$ for $j > 1$. 
To measure the sensitivity, we compare unpruned dynamics with those where the initially unoccupied SPFs are perturbed by adding higher order Krylov vectors: $\ket{\spf\kappa j}^\text{pert.} = \ket{\spf \kappa j} + r \ket{\spf\kappa{j+n_\kappa}}$, for $j>1$, where $r$ is a random number in the range $[-0.01,0.01]$. This procedure follows a suggestion by Meyer.\cite{meyer_communication_2017} As only initially unoccupied SPFs are perturbed, this procedure has no influence on the initial wavefunction.
The autocorrelation function is shown in  \fig\ref{fig:pyr24_A_acorr}, where the continuous black line corresponds to unpruned dynamics %
and the dashed red line corresponds to dynamics with perturbed initial SPFs.

This sensitivity to the choice of the initial SPFs renders a comparison of accuracy with the pruned dynamics difficult. In our pruned variant, SPFs that are dynamically added at later time steps, are prepared differently from the unpruned variant; see Section \ref{sec:impl_prune_A}. Even if the newly added SPFs were generated by the same procedure, the dynamics would differ because they would not be generated by the SPFs at $t=0$. 
To make the comparison with our pruned dynamics easier, we propagate until $t=t_S$ and restart the propagation again at $t=0$ with the subset of configurations from $t=t_S$. %
The SPFs at $t=0$ do not differ for the pruned and unpruned variants.
Note that this still corresponds to pruned dynamics for \emph{all} propagation times. It only means that we use a minimum pruned set of configurations for $t \le t_S$ but adding and removing other configurations is still allowed during this initial period.
We chose $t_S=\unit[7]{fs}$, well within the initial decay of the autocorrelation function. At least for dynamics with lower wave amplitude thresholds, during the first $\unit[7]{fs}$, almost all SPFs are used somewhere in $\tens A$ such that the number of initial SPFs are the same, compared to unpruned dynamics.
The full coefficient tensor is never required for the pruned dynamics.
Hence, this initialization procedure is negligible with respect to runtime because, compared to later times, not many configurations contribute to the wave function within the first $\unit[7]{fs}$.

In general, this procedure increases the stability of the dynamics. Only a single configuration is occupied at $t=0$ and many other configurations will become important within the first femtoseconds. Therefore, a dynamically pruned dynamics may require short time-steps during the initial period of propagation. 
For the pruning of the coefficient tensor, we use a ``configuration-space radius'' of $\sqrt{2}$,\cite{pvb_algorithms_tannor_2016} where $2D^2$ nearest neighbors are added for each non-negligible configuration. This further increased the stability, compared to the radius of $1$ used for pruning DVR,\cite{pW_tannor_2016} that only adds $2D$ nearest neighbors.

All computations were performed using a single core of Intel(R) Xeon(R) CPU E5-2650 v2 processors. The accuracy of the CMF propagator was set to $10^{-6}$ and the accuracy of the propagators for the \EOM of $\tens A$ and of the SPFs were set to $10^{-7}$. Like \lit{mctdh_selected_configurations_worth_2000}, we propagated until $\unit[80]{fs}$. %

\begin{figure}
 \includegraphics{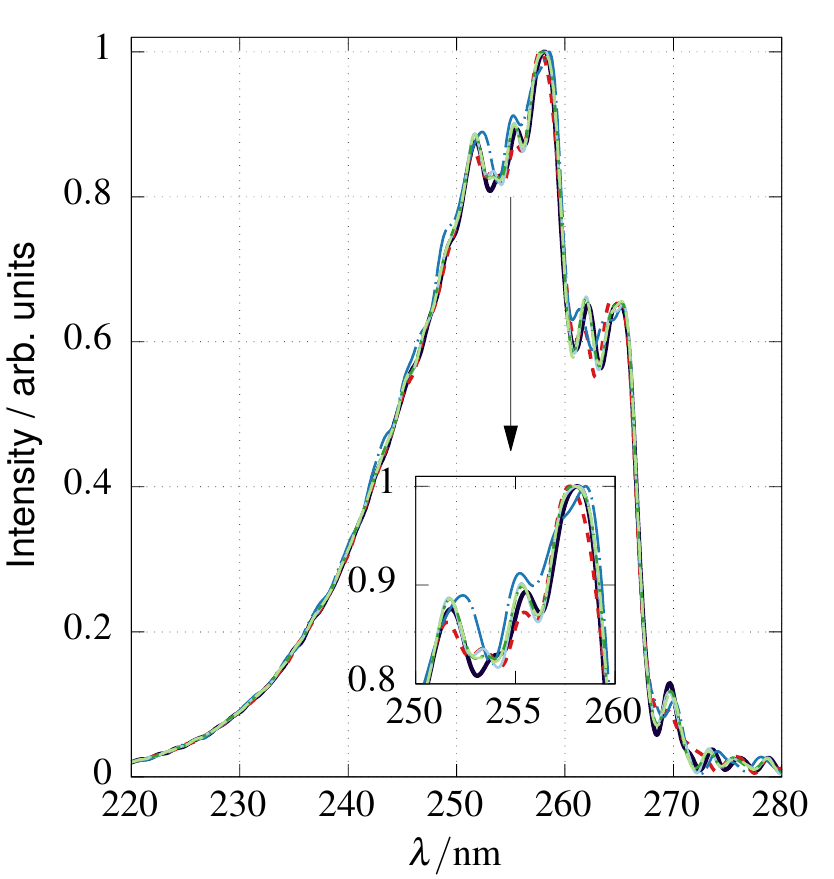}
 \caption{Shifted absorption spectrum of pyrazine using \mbox{(DP-)}MCTDH with variant A (\tab\ref{tab:pyr24_basis}) for the number of SPFs in each \DOF: unpruned (black line), unpruned with perturbed SPFs (red dashed line), pruned with an average usage of $0.7\%$ (blue, long dashed-dotted line), $4.4\%$ (pale blue, continuous line), $6.2\%$ (dashed-dotted green line) and $19\%$ (pale green, dashed line).
 See \fig\ref{fig:pyr24_A_acorr} and text for further details.}
 \label{fig:pyr24_A_spec}
\end{figure}

\Fig\ref{fig:pyr24_A_acorr} depicts the autocorrelation function for the pruned and unpruned dynamics (black line). Until approximately $\unit[80]{fs}$, the autocorrelation function can be well described using pruned dynamics. At later times, the autocorrelation function is still in good agreement, compared to the difference of unpruned dynamics with and without perturbed unoccupied SPFs (dashed red line), see above. In particular, the positions of the maxima of the autocorrelation function are generally in excellent agreement with those from the unpruned dynamics.
If $4.4\%$ of all possible configurations are used (pale blue line), the runtime can be reduced from 36 hours to approximately five hours. However, even if only $19\%$ of the configuration space is needed (dashed pale green line), the pruned dynamics takes $72\%$ of the runtime of the dynamics without pruning. This is caused by the additional overhead of the pruned tensor transformation. In general, pruning is more favorable if larger numbers of SPFs are required. Since, on average, only six SPFs in each degree of freedom are used, pruning cannot be fully efficient. Even so, a significant decrease in runtime is possible. 
We note that our pruned calculations achieve larger speed-ups than those found in the ML-spawning scheme.\cite{ml_spawning_gatti_2017} 
Furthermore, these timings are much better than those presented by Worth.\cite{mctdh_selected_configurations_worth_2000}. There, the static pruning led to \emph{larger} CPU time even  if only $16\%$ of the totally available configurations are used (compare configuration 24a3 to 24a in Table II in \lit{mctdh_selected_configurations_worth_2000}). Worth reported a speed-up of $4.6$ if only $0.5\%$ of configurations are used. We gain a speed-up of $5.1$ if $6.2\%$ of configurations are used and obtain a better converged spectrum, see below (\fig\ref{fig:pyr24_A_spec}).
Of course, these comparisons have to be taken with care, because the computations were done on different hardware. Wodraszka and Carrington used a pruning similar to that of Worth and developed a new algorithm that should improve performance.\cite{pruned_mctdh_carrington_2016} However, they did not try to solve the TDSE so no direct comparison can be done.

In many applications, not the autocorrelation function but the absorption spectrum is of importance. In general, this quantity is easier to converge than the autocorrelation function. We follow Worth\cite{mctdh_selected_configurations_worth_2000} and compute the spectrum as 
\begin{equation}
\begin{split}
 I(\omega) \propto \omega \int_0^T \dd t \Re\left[\braket{\Psi(0)}{\Psi(t)} \exp(\ii \omega t) \vphantom{\exp(-t/\tau)\cos\bigl(\pi t / (2T)\bigr)} \right.\\
 \left.\vphantom{\braket{\Psi(0)}{\Psi(t)} \exp(\ii \omega t)} \exp(-t/\tau)  \cos\bigl(\pi t / (2T)\bigr)\right],
 \end{split}
\end{equation}
with the damping parameter $\tau = \unit[150]{fs}$, using the \texttt{autospec} utility of the Heidelberg MCTDH package.\cite{mctdh_package} The shifted and scaled spectra of the autocorrelation functions of \fig\ref{fig:pyr24_A_acorr} are shown in \fig\ref{fig:pyr24_A_spec}. Compared to the spectrum of the unpruned dynamics with perturbed unoccupied SPFs, the agreement is excellent. Even using only $0.7\%$ of the full configuration space (blue line) gives a qualitatively correct behavior within the uncertainty of the spectrum introduced by the damping of the autocorrelation function ($\unit[1]{nm}$ at $\unit[250]{nm}$; see \lit{mctdh_selected_configurations_worth_2000}). If this accuracy is sufficient, the runtime can be reduced from 36 hours to 1 hour.
Note that the spectrum slightly differs from that shown in \lit{mctdh_selected_configurations_worth_2000} because we use the single-set approach and most likely a different accuracy of the propagator (no accuracy threshold is mentioned by Worth).

\subsubsection{Variant B (more SPFs)}
\label{subsec:pyr_varB}

\begin{figure*}
 \includegraphics{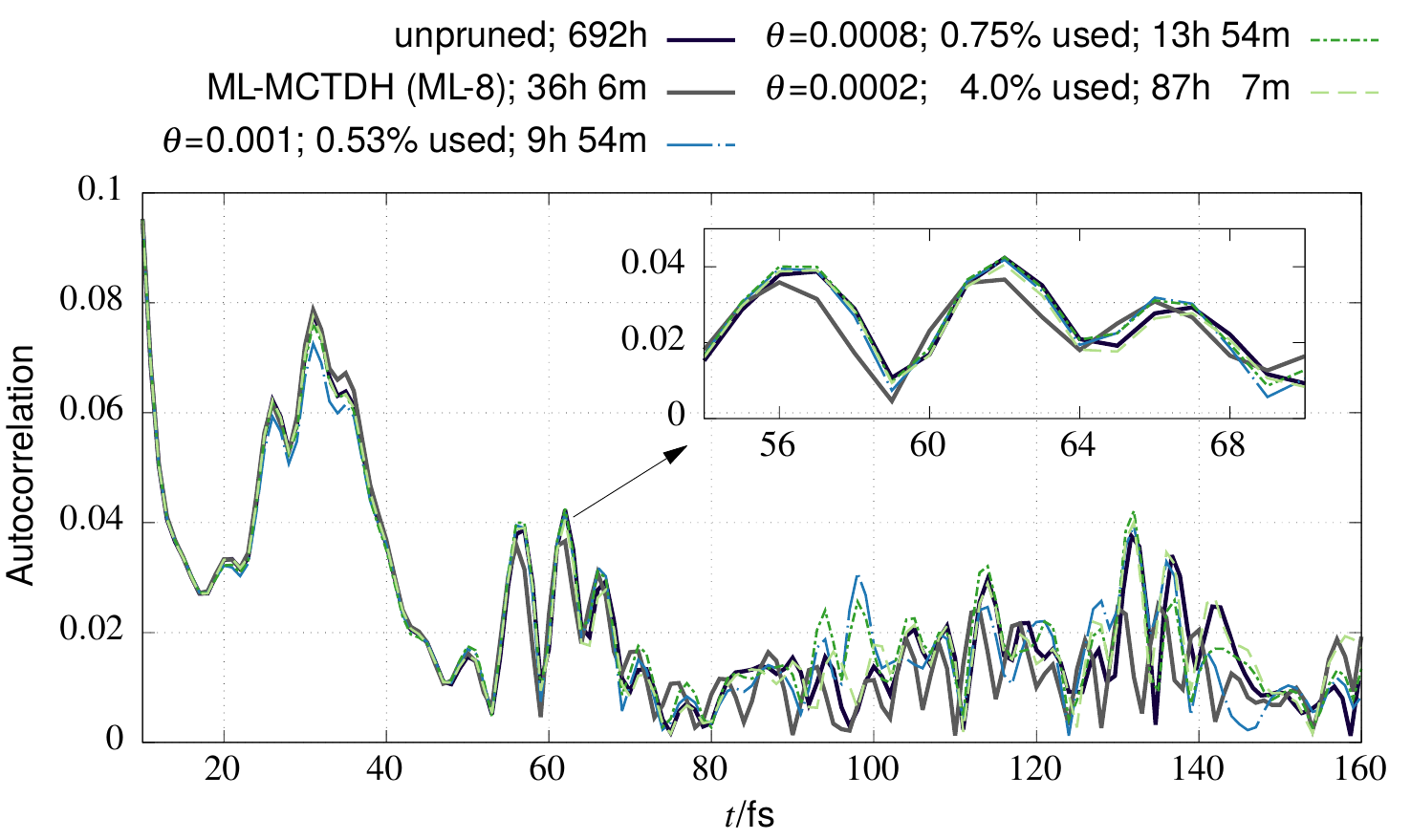}
 \caption{\newP{Same as \fig~\ref{fig:pyr24_A_acorr}, except that the maximal number of SPFs corresponds to variant B in \tab \ref{tab:pyr24_basis}.}
 The gray curve shows the result of ML-MCTDH (configuration ML-8 in \lit{ml_mctdh_meyer_2011}) \newP{for further comparisons}. 
 100\% corresponds to \newP{an average size of the pruned configuration tensor} of \new{$8.3\cdot 10^{7}$}.}
 \label{fig:pyr24_A_acorr_variantB}
\end{figure*}

We now turn to variant B where more SPFs in each \DOF are used. Now, the sensitivity to the initially unoccupied SPFs is not severe. The autocorrelation functions for pruned and unpruned dynamics are shown in \fig\ref{fig:pyr24_A_acorr_variantB}. Here, the unpruned MCTDH dynamics was performed with the Heidelberg package\cite{mctdh_package} in order to exploit  shared-memory parallelization (16 threads).\cite{mctdh_book_parallelization} Since the speed-up with respect to the number of threads is not ideal,\cite{ml_mctdh_meyer_2011} we have estimated the total runtime on a single core to be 692 hours (see \fig \ref{fig:pyr24_A_acorr_variantB}) based on a speed-up factor of $5.2$ for the parallel computation.
Until approximately $\unit[80]{fs}$, the autocorrelation function can be well described with pruned dynamics using only $0.53\%$ of the totally available configurations (blue curve). The runtime is then below 10 hours -- much less than even the runtime from unpruned dynamics of variant A! 
With only $0.75\%$ of all configurations (dashed-dotted green line), the autocorrelation function is accurately reproduced at almost all times. The deviation should be compared with that of the more accurate ML-MCTDH result (gray line); see next paragraph. The runtime was only about 14 hours, representing a speed-up of $49$.
Even if many more configurations are used, \eg $4\%$, (dashed pale green line), a significant speed-up of $7$ is obtained.

In \fig\ref{fig:pyr24_A_acorr_variantB}, we compare also to a ML-MCTDH calculation (gray curve). The set-up corresponds to that of the (largest) ``ML-8'' configuration in \lit{ml_mctdh_meyer_2011} but we use a lower accuracy of the propagator ($10^{-6}$), like we use in our other simulations. We employ the Heidelberg ML-MCTDH package\cite{mctdh_package_ml} for this simulation. 
This calculation is considered to be more accurate than variant B.\cite{ml_mctdh_meyer_2011} Despite the efficiency of the multilayer method, almost all of our pruned simulations are faster than the (unpruned) ML-MCTDH. Only if $4\%$ of the configurations are used is the pruned dynamics a factor of $2.4$ slower than the ML-8 simulations.
A simulation of the ``ML-6'' configuration from \lit{ml_mctdh_meyer_2011} (not shown) took only 14 hours and 46 minutes but this configuration is less accurate than variant B and still needs slightly more computing time than our more accurate pruned dynamics with usage of $0.75\%$, that took 13 hours and 54 minutes. Note, however, that Vendrell and Meyer comment that their ML-tree is not optimal.\cite{ml_mctdh_meyer_2011} 

\new{
The results of variants A and B are summarized in Table \ref{tab:dyn_comparison}. For evaluating the error, we follow Worth\cite{mctdh_selected_configurations_worth_2000} and evaluate the error as 
\begin{equation}
\Delta_{AR} = \| \vec{\tilde I}_A(\omega) - \vec{\tilde I}_R(\omega)\|_2 ,\quad \vec{\tilde I}_X = \vec I_X / \|\vec{I}_X\|_2,\label{eq:spectrum_error}
\end{equation}
where $R$ is the reference calculation. The norm is computed by discretizing the spectrum $I(\omega)$ on a grid of size $10^3$.
}

\begin{table}
\new{
 \caption{Comparison of the performance of variants A and B of DP-MCTDH dynamics of pyrazine with the ML variants from \lit{ml_mctdh_meyer_2011}. The errors are  defined according to \eq\eqref{eq:spectrum_error}. Error 1 uses the ML-8 calculation as a reference whereas error 2 uses the corresponding unpruned calculation as a reference. $\erw{N_\text{coeff}}$ denotes the (average) number of coefficients used to describe the configuration tensor.}
 \label{tab:dyn_comparison}
\begin{ruledtabular}
 \begin{tabular}{lcrcc}
 Set-up & $\erw{N_\text{coeff}}$ & runtime [h:m] & error 1 & error 2\\\hline
 ML-8 &  $1.5\cdot 10^5$ & 36:60 & - \\
 ML-7 &  $1.1 \cdot 10^5$ & 27:00 & 0.005 \\
 ML-6 &  $5.1\cdot 10^4$& 14:47 & 0.015 \\
 A: unpruned & $5.6\cdot 10^6$ & 36:10 & 0.044 & - \\
 A: -"-, perturbed & $5.6\cdot 10^6$ & 36:10 & 0.044 & 0.029\\
 A: $\theta=0.006$ & $3.9\cdot 10^4$ & 1:09 & 0.043 & 0.033\\
 A: $\theta=0.001$ & $2.5\cdot 10^5$ & 4:54 & 0.042 & 0.014\\
 A: $\theta=0.0008$ & $3.5\cdot 10^5$ & 7:16 & 0.043& 0.012\\
 A: $\theta=0.0002$ & $1.1\cdot 10^6$ & 26:09 & 0.045& 0.014\\
 B: unpruned & $8.3\cdot 10^7$ & 692:00& 0.011 & -\\
 B: $\theta=0.001$ & $4.4\cdot 10^5$ & 9:54 & 0.016 & 0.016\\
 B: $\theta=0.0008$ & $6.2\cdot 10^5$ & 13:54 & 0.013 & 0.011\\
 B: $\theta=0.0002$ & $3.3\cdot 10^6$ & 87:07 & 0.010 & 0.009\\
 \end{tabular}
 \end{ruledtabular}
}
\end{table}

\subsubsection{No restriction on the number of SPFs}
\label{subsec:pyr_noRestriction}

\begin{figure*}
 \includegraphics{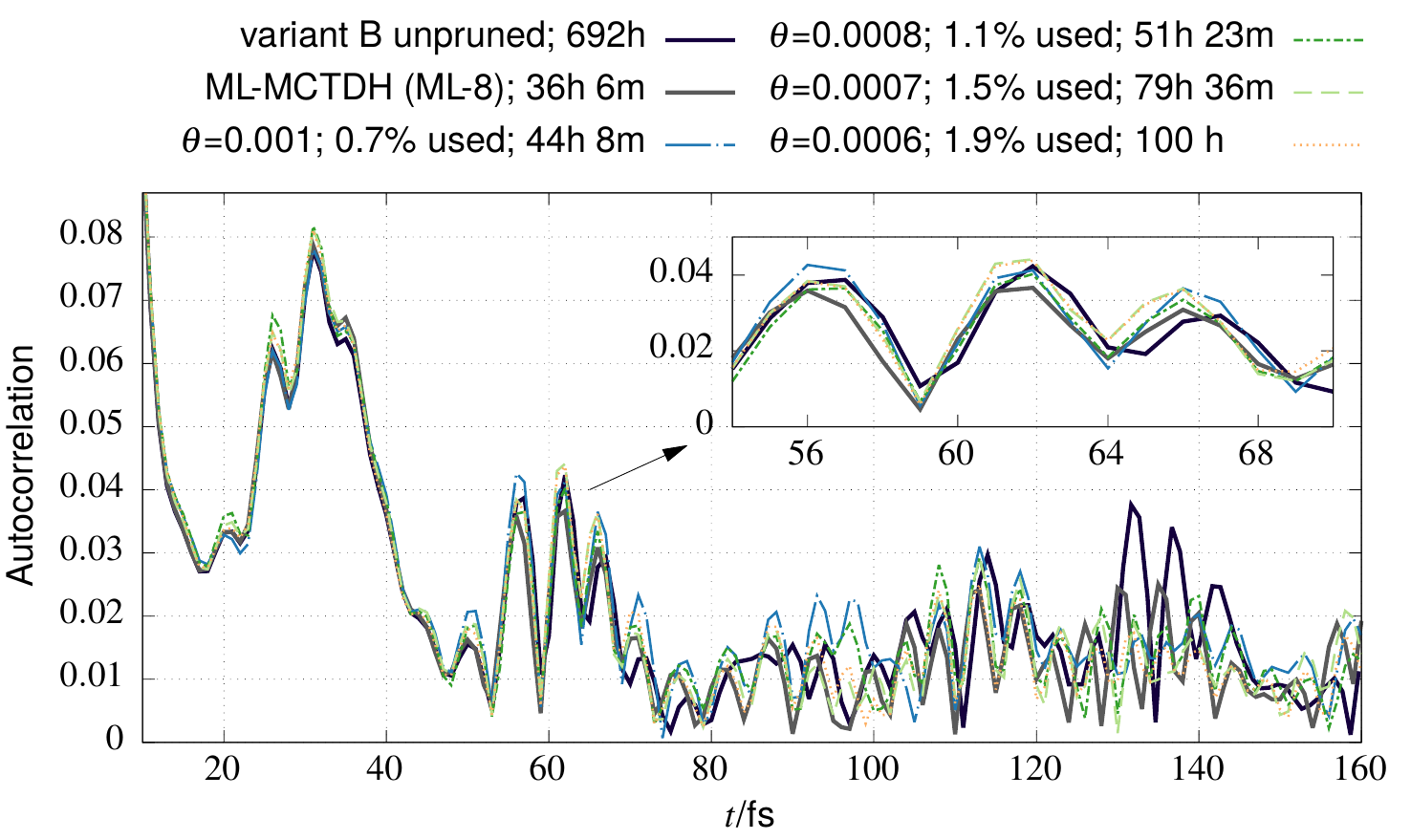}
 \caption{\newP{Same as \fig~\ref{fig:pyr24_A_acorr}, except that the maximal number of SPFs is as large as 50 in each dimension (except for the electronic \DOF) for the pruned dynamics. Except for dimension 1, this large number does not restrict the pruning. For further comparisons, the black line shows unpruned dynamics of variant B; see \tab \ref{tab:pyr24_basis}.}
 The gray curve shows result of ML-MCTDH (configuration ML-8 in \lit{ml_mctdh_meyer_2011}).
 100\% corresponds to \newP{an average size of the pruned configuration tensor} of $7.8\cdot 10^{13}$. 
 }
 \label{fig:pyr24_A_acorr_unconstraint}
\end{figure*}

As a final test, we lift the SPF constraint almost completely by setting the maximum number of SPFs in each \DOF to be 50, not including the electronic \DOF. 
The restriction to 50 SPFs was chosen for convenience. Except for dimension 1, this number does not restrict the pruning.
The resulting autocorrelation functions are shown in \fig\ref{fig:pyr24_A_acorr_unconstraint}. 
Comparing to the unpruned variant B and the more accurate ``ML-8'' variant, it shows that the autocorrelation function can be systematically converged within our pruning scheme. Notably, the positions of the maxima of the autocorrelation functions for the pruned dynamics (continuous blue, dashed(-dotted) green and dotted yellow lines) are generally different from variant B (black line) but resemble the more accurate ML-8 variant (gray line). Only after $\unit[120]{fs}$ one can notice a more significant deviation. At these times, even the ML-8 variant is not converged and one might speculate that our pruned dynamics are more accurate. To verify this, a more systematic convergence study would be required. This is beyond the scope of this work. %

Here, all of our pruned simulations need more runtime than the ML simulation.
However, the runtimes are of similar magnitude while including a substantially larger configuration space. As discussed above, by constraining the number of SPFs to a lower limit, faster runtimes may be obtained if some accuracy is sacrificed.

\subsection{Pruning both the primitive and the SPF basis with higher-dimensional mode combination: Pyrazine}
\label{sec:ex_combined}
As a proof of principle, we show that pruning both the SPF and the primitive basis can be combined. We show further that pruning the primitive basis allows for higher-dimensional mode combination. We use again the pyrazine model (variant A in \tab \ref{tab:pyr24_basis}) but now combine the modes $\nu_2, \nu_{6b} $ and $\nu_{8b}$ together with $\nu_4, \nu_5$ and $\nu_3$ using twelve SPFs. This decreases the dimension of the coefficient tensor by one but requires propagating six-dimensional SPFs. Additionally, this increases the overall runtime for  the unpruned propagation by 60\%. The propagation of the SPFs in this mode is then the dominant part (70\% of runtime) of the propagation.

\begin{figure*}
 \includegraphics{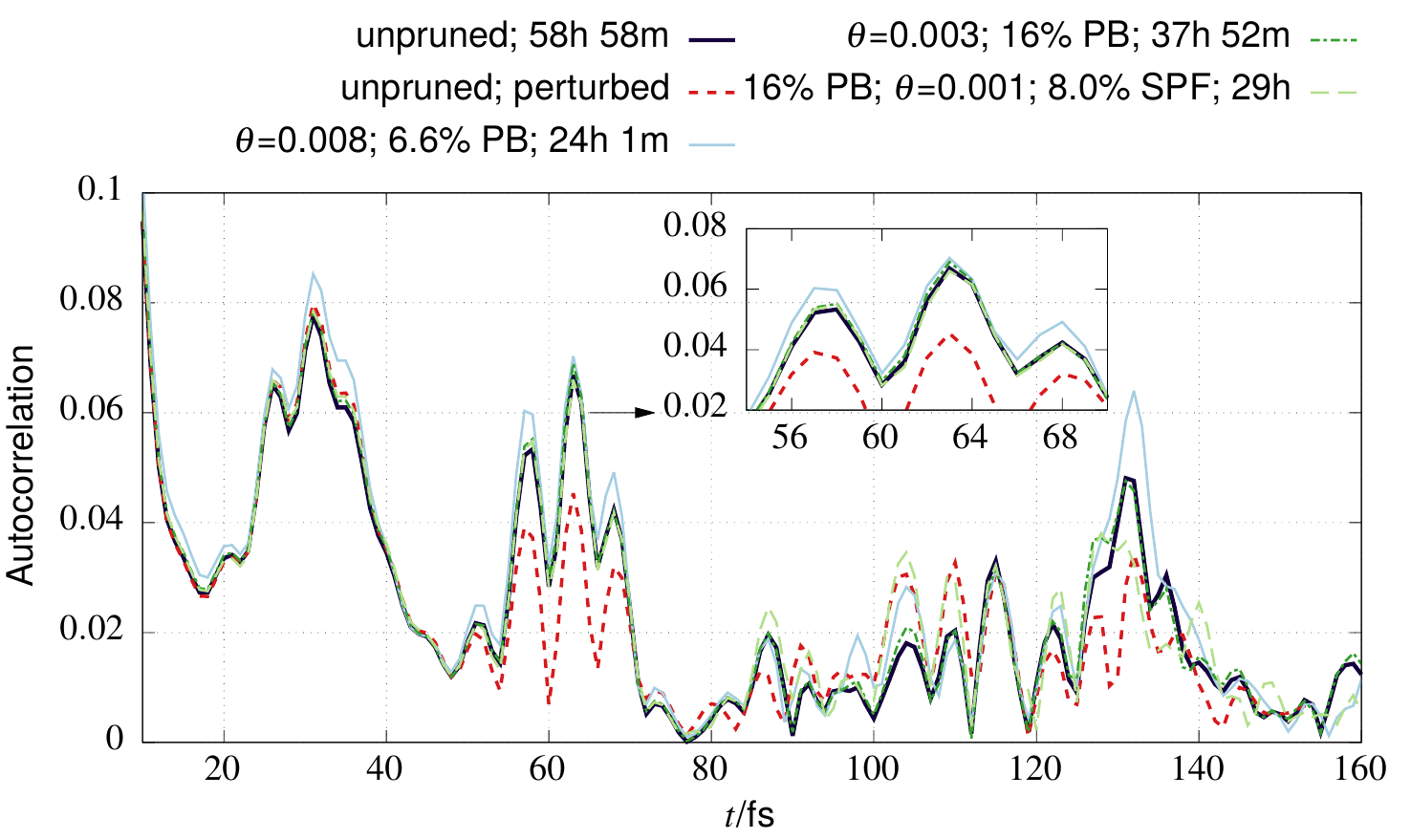}
 \caption{\newP{Same as \fig~\ref{fig:pyr24_A_acorr} but with higher-dimensional mode combination and with a pruned primitive basis (PB) in the first \DOF; see text.} 
 The percentages denote the average usage of \newP{PB}.
 100\% of \newP{PB} corresponds to a size of $1.2\cdot 10^{6}$. The dashed pale green line shows dynamics where both the primitive basis \newP{and the configuration space} are pruned. The SPFs span a space of size $1.3\cdot 10^6$.}
 \label{fig:pyr24_SS34}
\end{figure*}

Hence, it makes sense to prune the primitive basis. We prune only the DVR representation of these six-dimensional SPFs. Note that the Gauss-Hermite DVR gives a non-equidistant grid. Nevertheless, our pruning methodology from \lit{pW_tannor_2016} works here.
All other SPFs are propagated without pruning their representation. %
The result is shown in \fig\ref{fig:pyr24_SS34}. If only 16\% of the primitive basis is used (dashed-dotted green line in \fig\ref{fig:pyr24_SS34}), the autocorrelation function can be reproduced accurately and the runtime can be reduced from 59 hours to 38 hours. This almost matches the runtime of the unpruned dynamics from Section \ref{sec:ex_pyr}, thus offsetting the unfavorable mode combination set-up. %

If both the primitive basis and the SPF basis are pruned, the runtime can be decreased to 29 hours while retaining a reasonable accuracy (dashed pale green line).  Then, solving the SPF \EOM again becomes the dominant part of the runtime. The contributions of the parts of the code dealing with $\tens A$ is then less than 8\%.

\subsection{Analysis of the pruned configuration space}
\label{subsec:pyr_anal}
\begin{figure*}
\includegraphics{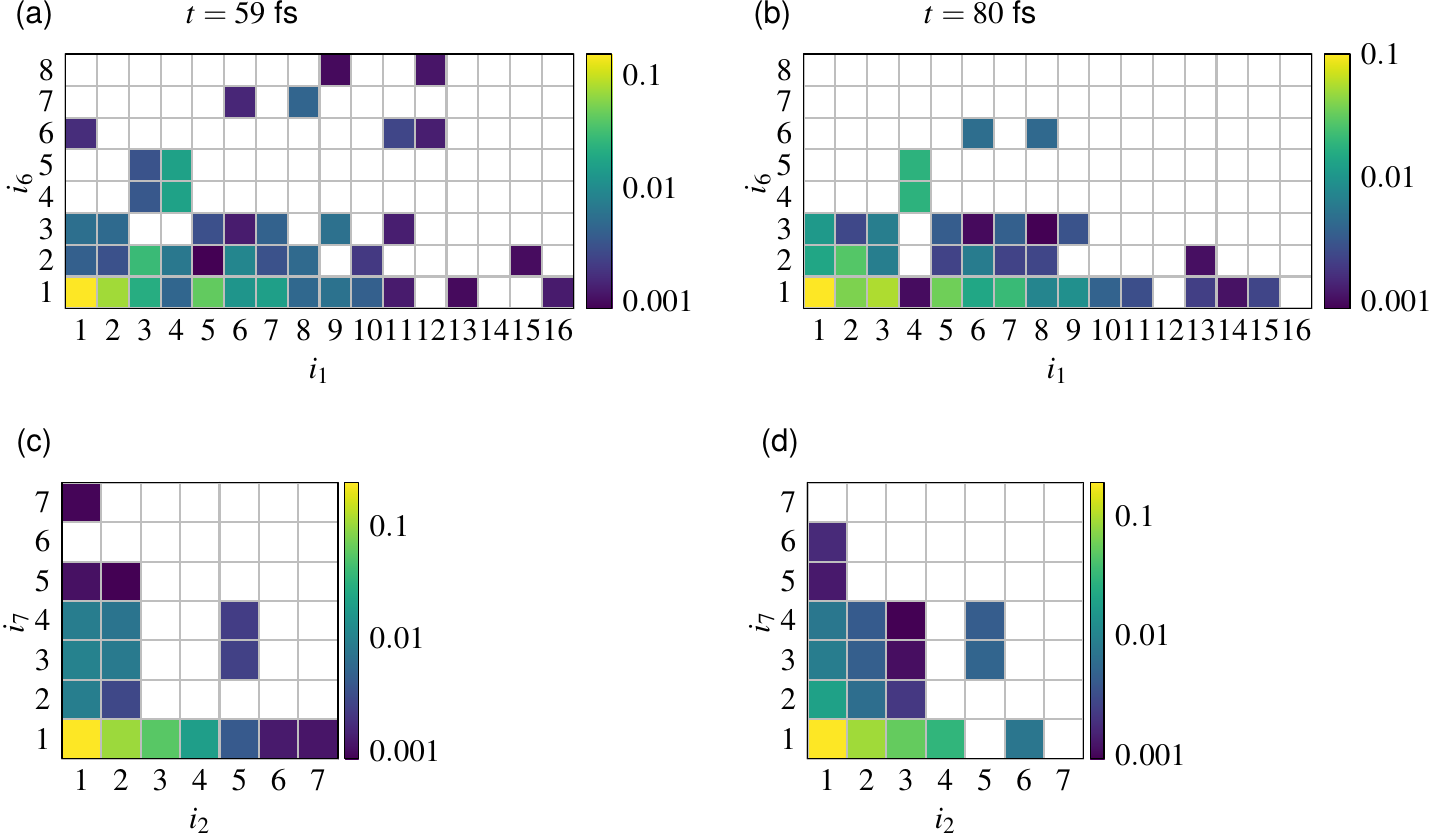}
    \caption{Graphical representation of the diagonal of the two-``particle'' reduced density matrix in configuration space of pyrazine at $t=\unit[59]{fs}$ for (a) and (c) and at $t=\unit[80]{fs}$ for (b) and (d). The density corresponds to unconstrained DP-MCTDH dynamics of \fig\ref{fig:pyr24_A_acorr_unconstraint} with a mean-usage of $1.1\%$. Each pixel corresponds to one configuration. White pixels correspond to configurations whose values are below the employed wave amplitude threshold, $0.0008$.
 Lower indices of the SPFs correspond to natural orbitals with larger populations. Panels (a) and (b) shows the (combined) \DOF 1 (modes $[\nu_{10a}$, $\nu_{6a}]$) and 6 ($[\nu_{19b}$,$\nu_{18b}]$); see \tab\ref{tab:pyr24_basis}. Panels (c) and (d) shows the \DOF 2 ($[\nu_{1}$,$\nu_{9a}$,$\nu_{8a}]$) and 7 ($[\nu_{18a}$, $\nu_{14}$, $\nu_{19a}$, $\nu_{17a}]$).
 }
 \label{fig:boxes}
\end{figure*}

We now turn to an analysis of the pruned configuration space used in Subsection \ref{subsec:pyr_noRestriction}.
It is clear that this nine-dimensional configuration space cannot be analyzed in detail everywhere.
It is instructive, however, to examine the diagonal of the two-``particle'' reduced density matrix in configuration space, \ie we have integrated the absolute square of the wavefunction in configuration space for all but two combined modes in the pruned subspace. One ``particle'' then corresponds to the combined modes used for setting up the SPFs in the corresponding \DOF. 
For example, the particle of the \DOF $6$ would represent the modes $\nu_{19b}$ and $\nu_{18b}$, compare with \tab\ref{tab:pyr24_basis}.
\Fig\ref{fig:boxes} shows two examples of the diagonal of the reduced density matrix of the underyling wavefunction in \fig\ref{fig:pyr24_A_acorr_unconstraint} with $1.1\%$ usage (dashed-dotted green curve) at $t=\unit[59]{fs}$ and $t=\unit[80]{fs}$. Since natural orbitals, ordered by their weight, are used for the dynamics, a triangle-shaped occupation of the ordered orbitals is expected. The smaller the orbital index, the larger the weight of the corresponding orbital.
This shape approximately appears in panels (c) and (d) in \fig\ref{fig:boxes}. However, the structure is more complex and contains some ``islands'' in this two-dimensional space. This is more pronounced in panels (a) and (b), showing \DOF $1$ and $6$. Because natural orbitals diagonalize only the one-particle density matrix, the natural populations (eigenvalues of the density matrix) are highly averaged quantities. It is still possible that configurations corresponding to orbitals with small natural populations may become significant during the dynamics. 
Note that the structure of these plots changes during time and can become more complex for intermediate times, as can be seen by comparing panels (a) with (b) or (c) with (d). Furthermore, note that the two-dimensional reduced densities vastly understate the sparsity of the full nine-dimensional density.
An analysis of the diagonal of the three-``particle'' density matrices forms a similar picture.
Additionally, an analysis of the DP-MCTDH dynamics with $0.7\%$ usage (dashed-dotted blue curve in \fig\ref{fig:pyr24_A_acorr_unconstraint}) and of unpruned dynamics of variant A does not significantly change the structure presented in \fig\ref{fig:boxes}.
Clearly, a simple static pruning as used in \lit{mctdh_selected_configurations_worth_2000,pruned_mctdh_carrington_2016} would be suboptimal as it would have to be vastly overextended to describe the structure of this non-direct-product configuration space for all simulation times.

\section{Conclusions and outlook}
\label{sec:conclusion}

We have presented two approaches for dynamical pruning MCTDH (DP-MCTDH). The first method prunes the primitive basis used to represent the SPFs whereas the second method prunes the set of configurations in the SPF space. 
The former method is useful for dynamics where a large, possibly multi-dimensional primitive basis set is required, as in \ce{NO2}. 
There, pruning the primitive bases leads to speed-up factors between two and three without jeopardizing accuracy.
We showed further that pruning the primitive basis makes higher-dimensional mode combination feasible, taking the 24-dimensional nonadiabatic dynamics of pyrazine as an example. Note that this partially relaxes the constraint that the Hamiltonian be of \SOP form.

Pruning the space of SPF configurations often allows for even larger time-savings. 
We again used pyrazine dynamics as an example and showed that, while retaining accuracy, pruning leads to speed-ups between $5$ and $50$.
The more SPFs are needed in MCTDH, the larger the speed-up in DP-MCTDH.
Depending on the setup, the runtime of the DP-MCTDH dynamics is comparable to or even faster than ML-MCTDH.
We also showed that both pruning methods can be combined, again using pyrazine with higher-dimensional mode combination as an example.

Pruning the SPF space dynamically introduces an MCTDH variant, DP-MCTDH, with just one single parameter, the wave amplitude threshold. 
By analyzing the reduced densities, we showed that dynamic instead of static pruning is crucial for a most effective reduction of the number of used configurations.
The result is a more powerful alternative to the ML-spawning scheme. 
Further, both pruning strategies might be useful for solving the TISE as well. There, an adaptive or iterative pruning like it is used in our time-independent simulations~\cite{pvb_algorithms_tannor_2016} should, in principle, be more efficient than simple static pruning\new{, as was recently shown.\cite{wodraszka_2017}}
The achievement of efficiency comparable to ML-MCTDH is promising. Combining pruning strategies with ML-MCTDH, giving DP-ML-MCTDH, should give an even faster method without jeopardizing accuracy.

\begin{acknowledgments}
We thank H.-D.~Meyer for providing us with the Heidelberg MCTDH package and for helpful discussions.
H.~R.~L.~thanks W.~Koch for helpful discussions and comments on the manuscript.
H.~R.~L.~acknowledges B.~Hartke for his continuous support, helpful discussions and comments on the manuscript.
H.~R.~L.~acknowledges support by the Fonds der Chemischen Industrie and the Studienstiftung des deutschen Volkes.
D.~J.~T.~acknowledges support from the Israel Science Foundation (1094/16) and the German-Israeli Foundation for Scientific Research and Development (GIF).
\end{acknowledgments}

\appendix*
\new{
\section{Pseudo-Code of the pruning procedure}
\label{app:pseudocode}
After each CMF step, the pruning proceeds as follows:
\begin{algorithmic}
\If {prune primitive basis}
  \State $\bullet$ update primitive basis
  \State $\bullet$ set newly added coefficients to zero
  \State $\bullet$ do Löwdin orthogonalization of SPFs 
  \State $\bullet$ update data structures for tensor transformation
\EndIf
\If {prune coefficient tensor $\tens A$}
  \State $\bullet$ update $\tens A$
  \State $\bullet$ set newly added coefficients of $\tens A$ to zero
  \State $\bullet$ set new SPFs to Krylov vectors of uncorrelated Hamiltonian 
  \State $\bullet$ do Gram-Schmidt orthogonalization of new SPFs
  \State $\bullet$ transform to natural orbitals
  \State $\bullet$ update data structures for tensor transformation
\EndIf
\end{algorithmic}
}


\begin{thebibliography}{116}%
\makeatletter
\providecommand \@ifxundefined [1]{%
 \@ifx{#1\undefined}
}%
\providecommand \@ifnum [1]{%
 \ifnum #1\expandafter \@firstoftwo
 \else \expandafter \@secondoftwo
 \fi
}%
\providecommand \@ifx [1]{%
 \ifx #1\expandafter \@firstoftwo
 \else \expandafter \@secondoftwo
 \fi
}%
\providecommand \natexlab [1]{#1}%
\providecommand \enquote  [1]{``#1''}%
\providecommand \bibnamefont  [1]{#1}%
\providecommand \bibfnamefont [1]{#1}%
\providecommand \citenamefont [1]{#1}%
\providecommand \href@noop [0]{\@secondoftwo}%
\providecommand \href [0]{\begingroup \@sanitize@url \@href}%
\providecommand \@href[1]{\@@startlink{#1}\@@href}%
\providecommand \@@href[1]{\endgroup#1\@@endlink}%
\providecommand \@sanitize@url [0]{\catcode `\\12\catcode `\$12\catcode
  `\&12\catcode `\#12\catcode `\^12\catcode `\_12\catcode `\%12\relax}%
\providecommand \@@startlink[1]{}%
\providecommand \@@endlink[0]{}%
\providecommand \url  [0]{\begingroup\@sanitize@url \@url }%
\providecommand \@url [1]{\endgroup\@href {#1}{\urlprefix }}%
\providecommand \urlprefix  [0]{URL }%
\providecommand \Eprint [0]{\href }%
\providecommand \doibase [0]{http://dx.doi.org/}%
\providecommand \selectlanguage [0]{\@gobble}%
\providecommand \bibinfo  [0]{\@secondoftwo}%
\providecommand \bibfield  [0]{\@secondoftwo}%
\providecommand \translation [1]{[#1]}%
\providecommand \BibitemOpen [0]{}%
\providecommand \bibitemStop [0]{}%
\providecommand \bibitemNoStop [0]{.\EOS\space}%
\providecommand \EOS [0]{\spacefactor3000\relax}%
\providecommand \BibitemShut  [1]{\csname bibitem#1\endcsname}%
\let\auto@bib@innerbib\@empty
%
\bibitem [{\citenamefont {Tannor}(2007)}]{tannor_book}%
  \BibitemOpen
  \bibfield  {author} {\bibinfo {author} {\bibfnamefont {D.~J.}\ \bibnamefont
  {Tannor}},\ }\href@noop {} {\emph {\bibinfo {title} {Introduction to Quantum
  Mechanics: A Time-Dependent Perspective}}},\ \bibinfo {edition} {1st}\ ed.\
  (\bibinfo  {publisher} {University Science Books},\ \bibinfo {year}
  {2007})\BibitemShut {NoStop}%
\bibitem [{\citenamefont {Chen}\ \emph {et~al.}(2016)\citenamefont {Chen},
  \citenamefont {Shao}, \citenamefont {Chen}, \citenamefont {Yang},\ and\
  \citenamefont {Zhang}}]{C2H_H2_zhang_2016}%
  \BibitemOpen
  \bibfield  {author} {\bibinfo {author} {\bibfnamefont {L.}~\bibnamefont
  {Chen}}, \bibinfo {author} {\bibfnamefont {K.}~\bibnamefont {Shao}}, \bibinfo
  {author} {\bibfnamefont {J.}~\bibnamefont {Chen}}, \bibinfo {author}
  {\bibfnamefont {M.}~\bibnamefont {Yang}}, \ and\ \bibinfo {author}
  {\bibfnamefont {D.~H.}\ \bibnamefont {Zhang}},\ }\href {\doibase
  10.1063/1.4948996} {\bibfield  {journal} {\bibinfo  {journal} {J. Chem.
  Phys.}\ }\textbf {\bibinfo {volume} {144}},\ \bibinfo {pages} {194309}
  (\bibinfo {year} {2016})}\BibitemShut {NoStop}%
\bibitem [{\citenamefont {Song}\ \emph {et~al.}(2014)\citenamefont {Song},
  \citenamefont {Li}, \citenamefont {Yang}, \citenamefont {Lu},\ and\
  \citenamefont {Guo}}]{H2_NH2_guo_2014}%
  \BibitemOpen
  \bibfield  {author} {\bibinfo {author} {\bibfnamefont {H.}~\bibnamefont
  {Song}}, \bibinfo {author} {\bibfnamefont {J.}~\bibnamefont {Li}}, \bibinfo
  {author} {\bibfnamefont {M.}~\bibnamefont {Yang}}, \bibinfo {author}
  {\bibfnamefont {Y.}~\bibnamefont {Lu}}, \ and\ \bibinfo {author}
  {\bibfnamefont {H.}~\bibnamefont {Guo}},\ }\href {\doibase
  10.1039/C4CP02227K} {\bibfield  {journal} {\bibinfo  {journal} {Phys. Chem.
  Chem. Phys.}\ }\textbf {\bibinfo {volume} {16}},\ \bibinfo {pages} {17770}
  (\bibinfo {year} {2014})}\BibitemShut {NoStop}%
\bibitem [{\citenamefont
  {Carrington~Jr.}(2017)}]{rovib_spectra_rev_carrington_2017}%
  \BibitemOpen
  \bibfield  {author} {\bibinfo {author} {\bibfnamefont {T.}~\bibnamefont
  {Carrington~Jr.}},\ }\href {\doibase 10.1063/1.4979117} {\bibfield  {journal}
  {\bibinfo  {journal} {J. Chem. Phys.}\ }\textbf {\bibinfo {volume} {146}},\
  \bibinfo {pages} {120902} (\bibinfo {year} {2017})}\BibitemShut {NoStop}%
\bibitem [{\citenamefont {Meyer}, \citenamefont {Manthe},\ and\ \citenamefont
  {Cederbaum}(1990)}]{mctdh_cederbaum_1990}%
  \BibitemOpen
  \bibfield  {author} {\bibinfo {author} {\bibfnamefont {H.-D.}\ \bibnamefont
  {Meyer}}, \bibinfo {author} {\bibfnamefont {U.}~\bibnamefont {Manthe}}, \
  and\ \bibinfo {author} {\bibfnamefont {L.}~\bibnamefont {Cederbaum}},\ }\href
  {\doibase 10.1016/0009-2614(90)87014-I} {\bibfield  {journal} {\bibinfo
  {journal} {Chem. Phys. Lett.}\ }\textbf {\bibinfo {volume} {165}},\ \bibinfo
  {pages} {73 } (\bibinfo {year} {1990})}\BibitemShut {NoStop}%
\bibitem [{\citenamefont {Manthe}, \citenamefont {Meyer},\ and\ \citenamefont
  {Cederbaum}(1992{\natexlab{a}})}]{mctdh_NOCl_cederbaum_1992}%
  \BibitemOpen
  \bibfield  {author} {\bibinfo {author} {\bibfnamefont {U.}~\bibnamefont
  {Manthe}}, \bibinfo {author} {\bibfnamefont {H.}~\bibnamefont {Meyer}}, \
  and\ \bibinfo {author} {\bibfnamefont {L.~S.}\ \bibnamefont {Cederbaum}},\
  }\href {\doibase 10.1063/1.463007} {\bibfield  {journal} {\bibinfo  {journal}
  {J. Chem. Phys.}\ }\textbf {\bibinfo {volume} {97}},\ \bibinfo {pages} {3199}
  (\bibinfo {year} {1992}{\natexlab{a}})}\BibitemShut {NoStop}%
\bibitem [{\citenamefont {Beck}\ \emph {et~al.}(2000)\citenamefont {Beck},
  \citenamefont {J{\"a}ckle}, \citenamefont {Worth},\ and\ \citenamefont
  {Meyer}}]{mctdh_rev_meyer_2000}%
  \BibitemOpen
  \bibfield  {author} {\bibinfo {author} {\bibfnamefont {M.~H.}\ \bibnamefont
  {Beck}}, \bibinfo {author} {\bibfnamefont {A.}~\bibnamefont {J{\"a}ckle}},
  \bibinfo {author} {\bibfnamefont {G.~A.}\ \bibnamefont {Worth}}, \ and\
  \bibinfo {author} {\bibfnamefont {H.-D.}\ \bibnamefont {Meyer}},\ }\href@noop
  {} {\bibfield  {journal} {\bibinfo  {journal} {Phys. Rep.}\ }\textbf
  {\bibinfo {volume} {324}},\ \bibinfo {pages} {1} (\bibinfo {year}
  {2000})}\BibitemShut {NoStop}%
\bibitem [{\citenamefont {Huarte-Larrañaga}\ and\ \citenamefont
  {Manthe}(2001)}]{CH4_H_manthe_2001}%
  \BibitemOpen
  \bibfield  {author} {\bibinfo {author} {\bibfnamefont {F.}~\bibnamefont
  {Huarte-Larrañaga}}\ and\ \bibinfo {author} {\bibfnamefont {U.}~\bibnamefont
  {Manthe}},\ }\href {\doibase 10.1021/jp003579w} {\bibfield  {journal}
  {\bibinfo  {journal} {J. Phys. Chem. A}\ }\textbf {\bibinfo {volume} {105}},\
  \bibinfo {pages} {2522} (\bibinfo {year} {2001})}\BibitemShut {NoStop}%
\bibitem [{\citenamefont {Worth}, \citenamefont {Meyer},\ and\ \citenamefont
  {Cederbaum}(1998)}]{pyrazine_24d_cederbaum_1998}%
  \BibitemOpen
  \bibfield  {author} {\bibinfo {author} {\bibfnamefont {G.~A.}\ \bibnamefont
  {Worth}}, \bibinfo {author} {\bibfnamefont {H.-D.}\ \bibnamefont {Meyer}}, \
  and\ \bibinfo {author} {\bibfnamefont {L.~S.}\ \bibnamefont {Cederbaum}},\
  }\href {\doibase 10.1063/1.476947} {\bibfield  {journal} {\bibinfo  {journal}
  {J. Chem. Phys.}\ }\textbf {\bibinfo {volume} {109}},\ \bibinfo {pages}
  {3518} (\bibinfo {year} {1998})}\BibitemShut {NoStop}%
\bibitem [{\citenamefont {Raab}\ \emph {et~al.}(1999)\citenamefont {Raab},
  \citenamefont {Worth}, \citenamefont {Meyer},\ and\ \citenamefont
  {Cederbaum}}]{pyrazine_24d_cederbaum_1999}%
  \BibitemOpen
  \bibfield  {author} {\bibinfo {author} {\bibfnamefont {A.}~\bibnamefont
  {Raab}}, \bibinfo {author} {\bibfnamefont {G.~A.}\ \bibnamefont {Worth}},
  \bibinfo {author} {\bibfnamefont {H.-D.}\ \bibnamefont {Meyer}}, \ and\
  \bibinfo {author} {\bibfnamefont {L.~S.}\ \bibnamefont {Cederbaum}},\ }\href
  {\doibase 10.1063/1.478061} {\bibfield  {journal} {\bibinfo  {journal} {J.
  Chem. Phys.}\ }\textbf {\bibinfo {volume} {110}},\ \bibinfo {pages} {936}
  (\bibinfo {year} {1999})}\BibitemShut {NoStop}%
\bibitem [{\citenamefont {Wang}\ and\ \citenamefont
  {Thoss}(2003)}]{ml_mctdh_thoss_2003}%
  \BibitemOpen
  \bibfield  {author} {\bibinfo {author} {\bibfnamefont {H.}~\bibnamefont
  {Wang}}\ and\ \bibinfo {author} {\bibfnamefont {M.}~\bibnamefont {Thoss}},\
  }\href {\doibase 10.1063/1.1580111} {\bibfield  {journal} {\bibinfo
  {journal} {J. Chem. Phys.}\ }\textbf {\bibinfo {volume} {119}},\ \bibinfo
  {pages} {1289} (\bibinfo {year} {2003})}\BibitemShut {NoStop}%
\bibitem [{\citenamefont {Manthe}(2008)}]{ml_mctdh_manthe_2008}%
  \BibitemOpen
  \bibfield  {author} {\bibinfo {author} {\bibfnamefont {U.}~\bibnamefont
  {Manthe}},\ }\href
  {http://scitation.aip.org/content/aip/journal/jcp/128/16/10.1063/1.2902982}
  {\bibfield  {journal} {\bibinfo  {journal} {J. Chem. Phys.}\ }\textbf
  {\bibinfo {volume} {128}},\ \bibinfo {eid} {164116} (\bibinfo {year}
  {2008})}\BibitemShut {NoStop}%
\bibitem [{\citenamefont {Vendrell}\ and\ \citenamefont
  {Meyer}(2011)}]{ml_mctdh_meyer_2011}%
  \BibitemOpen
  \bibfield  {author} {\bibinfo {author} {\bibfnamefont {O.}~\bibnamefont
  {Vendrell}}\ and\ \bibinfo {author} {\bibfnamefont {H.-D.}\ \bibnamefont
  {Meyer}},\ }\href {10.1063/1.3535541} {\bibfield  {journal} {\bibinfo
  {journal} {J. Chem. Phys.}\ }\textbf {\bibinfo {volume} {134}},\ \bibinfo
  {eid} {044135} (\bibinfo {year} {2011})}\BibitemShut {NoStop}%
\bibitem [{\citenamefont {Wang}(2015)}]{ml_mctdh_rev_wang_2015}%
  \BibitemOpen
  \bibfield  {author} {\bibinfo {author} {\bibfnamefont {H.}~\bibnamefont
  {Wang}},\ }\href {\doibase 10.1021/acs.jpca.5b03256} {\bibfield  {journal}
  {\bibinfo  {journal} {J. Phys. Chem. A}\ }\textbf {\bibinfo {volume} {119}},\
  \bibinfo {pages} {7951} (\bibinfo {year} {2015})}\BibitemShut {NoStop}%
\bibitem [{\citenamefont {Xie}, \citenamefont {Zheng},\ and\ \citenamefont
  {Lan}(2015)}]{ml_mctdh_anthracene_c60_lan_2015}%
  \BibitemOpen
  \bibfield  {author} {\bibinfo {author} {\bibfnamefont {Y.}~\bibnamefont
  {Xie}}, \bibinfo {author} {\bibfnamefont {J.}~\bibnamefont {Zheng}}, \ and\
  \bibinfo {author} {\bibfnamefont {Z.}~\bibnamefont {Lan}},\ }\href {\doibase
  10.1063/1.4909521} {\bibfield  {journal} {\bibinfo  {journal} {J. Chem.
  Phys.}\ }\textbf {\bibinfo {volume} {142}},\ \bibinfo {pages} {084706}
  (\bibinfo {year} {2015})}\BibitemShut {NoStop}%
\bibitem [{\citenamefont {Schulze}\ \emph {et~al.}(2016)\citenamefont
  {Schulze}, \citenamefont {Shibl}, \citenamefont {Al-Marri},\ and\
  \citenamefont {Kühn}}]{ml_mctdh_FMO_kuehn_2016}%
  \BibitemOpen
  \bibfield  {author} {\bibinfo {author} {\bibfnamefont {J.}~\bibnamefont
  {Schulze}}, \bibinfo {author} {\bibfnamefont {M.~F.}\ \bibnamefont {Shibl}},
  \bibinfo {author} {\bibfnamefont {M.~J.}\ \bibnamefont {Al-Marri}}, \ and\
  \bibinfo {author} {\bibfnamefont {O.}~\bibnamefont {Kühn}},\ }\href
  {\doibase 10.1063/1.4948563} {\bibfield  {journal} {\bibinfo  {journal} {J.
  Chem. Phys.}\ }\textbf {\bibinfo {volume} {144}},\ \bibinfo {pages} {185101}
  (\bibinfo {year} {2016})}\BibitemShut {NoStop}%
\bibitem [{\citenamefont {Welsch}\ and\ \citenamefont
  {Manthe}(2012)}]{ml_mctdh_H_CH4_manthe_2012}%
  \BibitemOpen
  \bibfield  {author} {\bibinfo {author} {\bibfnamefont {R.}~\bibnamefont
  {Welsch}}\ and\ \bibinfo {author} {\bibfnamefont {U.}~\bibnamefont
  {Manthe}},\ }\href
  {http://scitation.aip.org/content/aip/journal/jcp/137/24/10.1063/1.4772585}
  {\bibfield  {journal} {\bibinfo  {journal} {J. Chem. Phys.}\ }\textbf
  {\bibinfo {volume} {137}},\ \bibinfo {eid} {244106} (\bibinfo {year}
  {2012})}\BibitemShut {NoStop}%
\bibitem [{\citenamefont {Meng}\ and\ \citenamefont
  {Meyer}(2014)}]{mlmctdh_versus_mctdh_H2COO_meyer_2014}%
  \BibitemOpen
  \bibfield  {author} {\bibinfo {author} {\bibfnamefont {Q.}~\bibnamefont
  {Meng}}\ and\ \bibinfo {author} {\bibfnamefont {H.-D.}\ \bibnamefont
  {Meyer}},\ }\href {\doibase 10.1063/1.4896201} {\bibfield  {journal}
  {\bibinfo  {journal} {J. Chem. Phys.}\ }\textbf {\bibinfo {volume} {141}},\
  \bibinfo {pages} {124309} (\bibinfo {year} {2014})}\BibitemShut {NoStop}%
\bibitem [{\citenamefont {Mendive-Tapia}\ \emph {et~al.}(2017)\citenamefont
  {Mendive-Tapia}, \citenamefont {Firmino}, \citenamefont {Meyer},\ and\
  \citenamefont {Gatti}}]{ml_spawning_gatti_2017}%
  \BibitemOpen
  \bibfield  {author} {\bibinfo {author} {\bibfnamefont {D.}~\bibnamefont
  {Mendive-Tapia}}, \bibinfo {author} {\bibfnamefont {T.}~\bibnamefont
  {Firmino}}, \bibinfo {author} {\bibfnamefont {H.-D.}\ \bibnamefont {Meyer}},
  \ and\ \bibinfo {author} {\bibfnamefont {F.}~\bibnamefont {Gatti}},\ }\href
  {\doibase 10.1016/j.chemphys.2016.08.031} {\bibfield  {journal} {\bibinfo
  {journal} {Chem. Phys.}\ }\textbf {\bibinfo {volume} {482}},\ \bibinfo
  {pages} {113 } (\bibinfo {year} {2017})}\BibitemShut {NoStop}%
\bibitem [{\citenamefont {Jäckle}\ and\ \citenamefont
  {Meyer}(1996)}]{potfit_meyer_1996}%
  \BibitemOpen
  \bibfield  {author} {\bibinfo {author} {\bibfnamefont {A.}~\bibnamefont
  {Jäckle}}\ and\ \bibinfo {author} {\bibfnamefont {H.-D.}\ \bibnamefont
  {Meyer}},\ }\href {\doibase 10.1063/1.471513} {\bibfield  {journal} {\bibinfo
   {journal} {J. Chem. Phys.}\ }\textbf {\bibinfo {volume} {104}},\ \bibinfo
  {pages} {7974} (\bibinfo {year} {1996})}\BibitemShut {NoStop}%
\bibitem [{\citenamefont {Peláez}\ and\ \citenamefont
  {Meyer}(2013)}]{multigrid_potfit_meyer_2013}%
  \BibitemOpen
  \bibfield  {author} {\bibinfo {author} {\bibfnamefont {D.}~\bibnamefont
  {Peláez}}\ and\ \bibinfo {author} {\bibfnamefont {H.-D.}\ \bibnamefont
  {Meyer}},\ }\href {\doibase 10.1063/1.4773021} {\bibfield  {journal}
  {\bibinfo  {journal} {J. Chem. Phys.}\ }\textbf {\bibinfo {volume} {138}},\
  \bibinfo {pages} {014108} (\bibinfo {year} {2013})}\BibitemShut {NoStop}%
\bibitem [{\citenamefont {Otto}(2014)}]{mctdh_multilayer_potfit_otto_2014}%
  \BibitemOpen
  \bibfield  {author} {\bibinfo {author} {\bibfnamefont {F.}~\bibnamefont
  {Otto}},\ }\href {\doibase 10.1063/1.4856135} {\bibfield  {journal} {\bibinfo
   {journal} {J. Chem. Phys.}\ }\textbf {\bibinfo {volume} {140}},\ \bibinfo
  {pages} {014106} (\bibinfo {year} {2014})}\BibitemShut {NoStop}%
\bibitem [{\citenamefont {Manzhos}\ and\ \citenamefont
  {Carrington}(2006)}]{pes_neural_network_sum_of_products_carrington_2006}%
  \BibitemOpen
  \bibfield  {author} {\bibinfo {author} {\bibfnamefont {S.}~\bibnamefont
  {Manzhos}}\ and\ \bibinfo {author} {\bibfnamefont {T.}~\bibnamefont
  {Carrington}},\ }\href {\doibase 10.1063/1.2387950} {\bibfield  {journal}
  {\bibinfo  {journal} {J. Chem. Phys.}\ }\textbf {\bibinfo {volume} {125}},\
  \bibinfo {pages} {194105} (\bibinfo {year} {2006})}\BibitemShut {NoStop}%
\bibitem [{\citenamefont {Koch}\ and\ \citenamefont
  {Zhang}(2014)}]{sop_pes_neural_networks_zhang_2014}%
  \BibitemOpen
  \bibfield  {author} {\bibinfo {author} {\bibfnamefont {W.}~\bibnamefont
  {Koch}}\ and\ \bibinfo {author} {\bibfnamefont {D.~H.}\ \bibnamefont
  {Zhang}},\ }\href {\doibase 10.1063/1.4887508} {\bibfield  {journal}
  {\bibinfo  {journal} {J. Chem. Phys.}\ }\textbf {\bibinfo {volume} {141}},\
  \bibinfo {pages} {021101} (\bibinfo {year} {2014})}\BibitemShut {NoStop}%
\bibitem [{\citenamefont {Avila}\ and\ \citenamefont
  {Carrington}(2015)}]{pes_sum_of_products_smolyak_carrington_2015}%
  \BibitemOpen
  \bibfield  {author} {\bibinfo {author} {\bibfnamefont {G.}~\bibnamefont
  {Avila}}\ and\ \bibinfo {author} {\bibfnamefont {T.}~\bibnamefont
  {Carrington}},\ }\href {\doibase 10.1063/1.4926651} {\bibfield  {journal}
  {\bibinfo  {journal} {J. Chem. Phys.}\ }\textbf {\bibinfo {volume} {143}},\
  \bibinfo {pages} {044106} (\bibinfo {year} {2015})}\BibitemShut {NoStop}%
\bibitem [{\citenamefont {Ziegler}\ and\ \citenamefont
  {Rauhut}(2016)}]{pes_sop_from_multimode_fit_rauhut_2016}%
  \BibitemOpen
  \bibfield  {author} {\bibinfo {author} {\bibfnamefont {B.}~\bibnamefont
  {Ziegler}}\ and\ \bibinfo {author} {\bibfnamefont {G.}~\bibnamefont
  {Rauhut}},\ }\href {\doibase 10.1063/1.4943985} {\bibfield  {journal}
  {\bibinfo  {journal} {J. Chem. Phys.}\ }\textbf {\bibinfo {volume} {144}},\
  \bibinfo {pages} {114114} (\bibinfo {year} {2016})}\BibitemShut {NoStop}%
\bibitem [{\citenamefont {Vendrell}\ \emph {et~al.}(2007)\citenamefont
  {Vendrell}, \citenamefont {Gatti}, \citenamefont {Lauvergnat},\ and\
  \citenamefont {Meyer}}]{zundel_mctdh_hamiltonian_meyer_2007}%
  \BibitemOpen
  \bibfield  {author} {\bibinfo {author} {\bibfnamefont {O.}~\bibnamefont
  {Vendrell}}, \bibinfo {author} {\bibfnamefont {F.}~\bibnamefont {Gatti}},
  \bibinfo {author} {\bibfnamefont {D.}~\bibnamefont {Lauvergnat}}, \ and\
  \bibinfo {author} {\bibfnamefont {H.-D.}\ \bibnamefont {Meyer}},\ }\href
  {\doibase 10.1063/1.2787588} {\bibfield  {journal} {\bibinfo  {journal} {J.
  Chem. Phys.}\ }\textbf {\bibinfo {volume} {127}},\ \bibinfo {pages} {184302}
  (\bibinfo {year} {2007})}\BibitemShut {NoStop}%
\bibitem [{\citenamefont {Manthe}(1996)}]{cdvr_manthe_1996}%
  \BibitemOpen
  \bibfield  {author} {\bibinfo {author} {\bibfnamefont {U.}~\bibnamefont
  {Manthe}},\ }\href {\doibase 10.1063/1.471847} {\bibfield  {journal}
  {\bibinfo  {journal} {J. Chem. Phys.}\ }\textbf {\bibinfo {volume} {105}},\
  \bibinfo {pages} {6989} (\bibinfo {year} {1996})}\BibitemShut {NoStop}%
\bibitem [{\citenamefont {Manthe}(2009)}]{ml_cdvr_2009}%
  \BibitemOpen
  \bibfield  {author} {\bibinfo {author} {\bibfnamefont {U.}~\bibnamefont
  {Manthe}},\ }\href {\doibase 10.1063/1.3069655} {\bibfield  {journal}
  {\bibinfo  {journal} {J. Chem. Phys.}\ }\textbf {\bibinfo {volume} {130}},\
  \bibinfo {pages} {054109} (\bibinfo {year} {2009})}\BibitemShut {NoStop}%
\bibitem [{\citenamefont {Manthe}(2015)}]{mctdh_denmat_inv_manthe_2015}%
  \BibitemOpen
  \bibfield  {author} {\bibinfo {author} {\bibfnamefont {U.}~\bibnamefont
  {Manthe}},\ }\href {\doibase 10.1063/1.4922889} {\bibfield  {journal}
  {\bibinfo  {journal} {J. Chem. Phys.}\ }\textbf {\bibinfo {volume} {142}},\
  \bibinfo {pages} {244109} (\bibinfo {year} {2015})}\BibitemShut {NoStop}%
\bibitem [{\citenamefont {Hartke}(2006)}]{proDG_hartke_2006}%
  \BibitemOpen
  \bibfield  {author} {\bibinfo {author} {\bibfnamefont {B.}~\bibnamefont
  {Hartke}},\ }\href {\doibase 10.1039/b606376d} {\bibfield  {journal}
  {\bibinfo  {journal} {Phys. Chem. Chem. Phys.}\ }\textbf {\bibinfo {volume}
  {8}},\ \bibinfo {pages} {3627} (\bibinfo {year} {2006})}\BibitemShut
  {NoStop}%
\bibitem [{\citenamefont {McCormack}(2006)}]{pruning_mccormack_2006}%
  \BibitemOpen
  \bibfield  {author} {\bibinfo {author} {\bibfnamefont {D.~A.}\ \bibnamefont
  {McCormack}},\ }\href {\doibase 10.1063/1.2196889} {\bibfield  {journal}
  {\bibinfo  {journal} {J. Chem. Phys.}\ }\textbf {\bibinfo {volume} {124}},\
  \bibinfo {pages} {204101} (\bibinfo {year} {2006})}\BibitemShut {NoStop}%
\bibitem [{\citenamefont {Pettey}\ and\ \citenamefont
  {Wyatt}(2006)}]{pruning_wyatt_2006}%
  \BibitemOpen
  \bibfield  {author} {\bibinfo {author} {\bibfnamefont {L.~R.}\ \bibnamefont
  {Pettey}}\ and\ \bibinfo {author} {\bibfnamefont {R.~E.}\ \bibnamefont
  {Wyatt}},\ }\href {\doibase 10.1016/j.cplett.2006.04.081} {\bibfield
  {journal} {\bibinfo  {journal} {Chem. Phys. Lett.}\ }\textbf {\bibinfo
  {volume} {424}},\ \bibinfo {pages} {443 } (\bibinfo {year}
  {2006})}\BibitemShut {NoStop}%
\bibitem [{\citenamefont {Pettey}\ and\ \citenamefont
  {Wyatt}(2007)}]{pruning_wyatt_2007}%
  \BibitemOpen
  \bibfield  {author} {\bibinfo {author} {\bibfnamefont {L.~R.}\ \bibnamefont
  {Pettey}}\ and\ \bibinfo {author} {\bibfnamefont {R.~E.}\ \bibnamefont
  {Wyatt}},\ }\href {\doibase 10.1002/qua.21301} {\bibfield  {journal}
  {\bibinfo  {journal} {Int. J. Quantum Chem.}\ }\textbf {\bibinfo {volume}
  {107}},\ \bibinfo {pages} {1566} (\bibinfo {year} {2007})}\BibitemShut
  {NoStop}%
\bibitem [{\citenamefont {Takemoto}, \citenamefont {Shimshovitz},\ and\
  \citenamefont {Tannor}(2012)}]{pvb_edyn_takemoto_tannor_2012}%
  \BibitemOpen
  \bibfield  {author} {\bibinfo {author} {\bibfnamefont {N.}~\bibnamefont
  {Takemoto}}, \bibinfo {author} {\bibfnamefont {A.}~\bibnamefont
  {Shimshovitz}}, \ and\ \bibinfo {author} {\bibfnamefont {D.~J.}\ \bibnamefont
  {Tannor}},\ }\href {\doibase 10.1063/1.4732306} {\bibfield  {journal}
  {\bibinfo  {journal} {J. Chem. Phys.}\ }\textbf {\bibinfo {volume} {137}},\
  \bibinfo {pages} {011102} (\bibinfo {year} {2012})}\BibitemShut {NoStop}%
\bibitem [{\citenamefont {Assémat}, \citenamefont {Machnes},\ and\
  \citenamefont {Tannor}(2015)}]{pvb_edyn_tannor_2015}%
  \BibitemOpen
  \bibfield  {author} {\bibinfo {author} {\bibfnamefont {E.}~\bibnamefont
  {Assémat}}, \bibinfo {author} {\bibfnamefont {S.}~\bibnamefont {Machnes}}, \
  and\ \bibinfo {author} {\bibfnamefont {D.}~\bibnamefont {Tannor}},\
  }\href@noop {} {\enquote {\bibinfo {title} {Double ionization of {Helium}
  from a phase space perspective},}\ } (\bibinfo {year} {2015}),\ \Eprint
  {http://arxiv.org/abs/arXiv:1502.05165} {arXiv:1502.05165} \BibitemShut
  {NoStop}%
\bibitem [{\citenamefont {Larsson}, \citenamefont {Hartke},\ and\ \citenamefont
  {Tannor}(2016)}]{pW_tannor_2016}%
  \BibitemOpen
  \bibfield  {author} {\bibinfo {author} {\bibfnamefont {H.~R.}\ \bibnamefont
  {Larsson}}, \bibinfo {author} {\bibfnamefont {B.}~\bibnamefont {Hartke}}, \
  and\ \bibinfo {author} {\bibfnamefont {D.~J.}\ \bibnamefont {Tannor}},\
  }\href {\doibase 10.1063/1.4967432} {\bibfield  {journal} {\bibinfo
  {journal} {J. Chem. Phys.}\ }\textbf {\bibinfo {volume} {145}},\ \bibinfo
  {pages} {204108} (\bibinfo {year} {2016})}\BibitemShut {NoStop}%
\bibitem [{\citenamefont {Light}(1992)}]{dvr_rev_light_1992}%
  \BibitemOpen
  \bibfield  {author} {\bibinfo {author} {\bibfnamefont {J.~C.}\ \bibnamefont
  {Light}},\ }\enquote {\bibinfo {title} {{Discrete Variable Representations in
  Quantum Dynamics}},}\ in\ \href@noop {} {\emph {\bibinfo {booktitle}
  {Time-Dependent Quantum Molecular Dynamics}}},\ \bibinfo {editor} {edited by\
  \bibinfo {editor} {\bibfnamefont {J.}~\bibnamefont {Broeckhove}}\ and\
  \bibinfo {editor} {\bibfnamefont {L.}~\bibnamefont {Lathouwers}}}\ (\bibinfo
  {publisher} {Springer},\ \bibinfo {year} {1992})\ pp.\ \bibinfo {pages}
  {185--199}\BibitemShut {NoStop}%
\bibitem [{\citenamefont {Davis}\ and\ \citenamefont
  {Heller}(1979)}]{semicl_gauss_heller_1979}%
  \BibitemOpen
  \bibfield  {author} {\bibinfo {author} {\bibfnamefont {M.~J.}\ \bibnamefont
  {Davis}}\ and\ \bibinfo {author} {\bibfnamefont {E.~J.}\ \bibnamefont
  {Heller}},\ }\href {\doibase 10.1063/1.438727} {\bibfield  {journal}
  {\bibinfo  {journal} {J. Chem. Phys.}\ }\textbf {\bibinfo {volume} {71}},\
  \bibinfo {pages} {3383} (\bibinfo {year} {1979})}\BibitemShut {NoStop}%
\bibitem [{\citenamefont {Shimshovitz}\ and\ \citenamefont
  {Tannor}(2012)}]{pvb_tannor_2012}%
  \BibitemOpen
  \bibfield  {author} {\bibinfo {author} {\bibfnamefont {A.}~\bibnamefont
  {Shimshovitz}}\ and\ \bibinfo {author} {\bibfnamefont {D.~J.}\ \bibnamefont
  {Tannor}},\ }\href {\doibase 10.1103/physrevlett.109.070402} {\bibfield
  {journal} {\bibinfo  {journal} {Phys. Rev. Lett.}\ }\textbf {\bibinfo
  {volume} {109}},\ \bibinfo {pages} {070402} (\bibinfo {year}
  {2012})}\BibitemShut {NoStop}%
\bibitem [{\citenamefont {Machnes}\ \emph {et~al.}(2016)\citenamefont
  {Machnes}, \citenamefont {Assémat}, \citenamefont {Larsson},\ and\
  \citenamefont {Tannor}}]{pvb_math_tannor_2016}%
  \BibitemOpen
  \bibfield  {author} {\bibinfo {author} {\bibfnamefont {S.}~\bibnamefont
  {Machnes}}, \bibinfo {author} {\bibfnamefont {E.}~\bibnamefont {Assémat}},
  \bibinfo {author} {\bibfnamefont {H.~R.}\ \bibnamefont {Larsson}}, \ and\
  \bibinfo {author} {\bibfnamefont {D.~J.}\ \bibnamefont {Tannor}},\ }\href
  {\doibase 10.1021/acs.jpca.5b12370} {\bibfield  {journal} {\bibinfo
  {journal} {J. Phys. Chem. A}\ }\textbf {\bibinfo {volume} {120}},\ \bibinfo
  {pages} {3296} (\bibinfo {year} {2016})}\BibitemShut {NoStop}%
\bibitem [{\citenamefont {Tannor}\ \emph {et~al.}(2017)\citenamefont {Tannor},
  \citenamefont {Machnes}, \citenamefont {Assémat},\ and\ \citenamefont
  {Larsson}}]{pvb_rev_tannor_2017}%
  \BibitemOpen
  \bibfield  {author} {\bibinfo {author} {\bibfnamefont {D.~J.}\ \bibnamefont
  {Tannor}}, \bibinfo {author} {\bibfnamefont {S.}~\bibnamefont {Machnes}},
  \bibinfo {author} {\bibfnamefont {E.}~\bibnamefont {Assémat}}, \ and\
  \bibinfo {author} {\bibfnamefont {H.~R.}\ \bibnamefont {Larsson}},\ }\enquote
  {\bibinfo {title} {Phase space vs.~coordinate space methods: Prognosis for
  large quantum calculations},}\ in\ \href@noop {} {\emph {\bibinfo {booktitle}
  {Advances in Chemical Physics}}},\ Vol.\ \bibinfo {volume} {163}\ (\bibinfo
  {publisher} {John Wiley \& Sons, Inc.},\ \bibinfo {year} {2017})\ p.\
  \bibinfo {pages} {in press}\BibitemShut {NoStop}%
\bibitem [{\citenamefont {Wilson}(1987)}]{gabor_basis_wilson_1987}%
  \BibitemOpen
  \bibfield  {author} {\bibinfo {author} {\bibfnamefont {K.~G.}\ \bibnamefont
  {Wilson}},\ }\href@noop {} {\enquote {\bibinfo {title} {Generalized wannier
  functions},}\ }\bibinfo {howpublished} {Cornell University preprint}
  (\bibinfo {year} {1987})\BibitemShut {NoStop}%
\bibitem [{\citenamefont {Daubechies}, \citenamefont {Jaffard},\ and\
  \citenamefont {Journé}(1991)}]{wilson_gabor_basis_journe_1991}%
  \BibitemOpen
  \bibfield  {author} {\bibinfo {author} {\bibfnamefont {I.}~\bibnamefont
  {Daubechies}}, \bibinfo {author} {\bibfnamefont {S.}~\bibnamefont {Jaffard}},
  \ and\ \bibinfo {author} {\bibfnamefont {J.-L.}\ \bibnamefont {Journé}},\
  }\href {\doibase 10.1137/0522035} {\bibfield  {journal} {\bibinfo  {journal}
  {SIAM J. Math. Anal.}\ }\textbf {\bibinfo {volume} {22}},\ \bibinfo {pages}
  {554} (\bibinfo {year} {1991})}\BibitemShut {NoStop}%
\bibitem [{\citenamefont {Poirier}(2003)}]{weylet_1_poirier_2003}%
  \BibitemOpen
  \bibfield  {author} {\bibinfo {author} {\bibfnamefont {B.}~\bibnamefont
  {Poirier}},\ }\href {\doibase 10.1142/S0219633603000380} {\bibfield
  {journal} {\bibinfo  {journal} {J. Theor. Comp. Chem.}\ }\textbf {\bibinfo
  {volume} {02}},\ \bibinfo {pages} {65} (\bibinfo {year} {2003})}\BibitemShut
  {NoStop}%
\bibitem [{\citenamefont {Poirier}\ and\ \citenamefont
  {Salam}(2004{\natexlab{a}})}]{weylet_2_poirier_2004}%
  \BibitemOpen
  \bibfield  {author} {\bibinfo {author} {\bibfnamefont {B.}~\bibnamefont
  {Poirier}}\ and\ \bibinfo {author} {\bibfnamefont {A.}~\bibnamefont
  {Salam}},\ }\href {\doibase 10.1063/1.1767511} {\bibfield  {journal}
  {\bibinfo  {journal} {J. Chem. Phys.}\ }\textbf {\bibinfo {volume} {121}},\
  \bibinfo {pages} {1690} (\bibinfo {year} {2004}{\natexlab{a}})}\BibitemShut
  {NoStop}%
\bibitem [{\citenamefont {Poirier}\ and\ \citenamefont
  {Salam}(2004{\natexlab{b}})}]{weylet_3_poirier_2004}%
  \BibitemOpen
  \bibfield  {author} {\bibinfo {author} {\bibfnamefont {B.}~\bibnamefont
  {Poirier}}\ and\ \bibinfo {author} {\bibfnamefont {A.}~\bibnamefont
  {Salam}},\ }\href {\doibase 10.1063/1.1767512} {\bibfield  {journal}
  {\bibinfo  {journal} {J. Chem. Phys.}\ }\textbf {\bibinfo {volume} {121}},\
  \bibinfo {pages} {1704} (\bibinfo {year} {2004}{\natexlab{b}})}\BibitemShut
  {NoStop}%
\bibitem [{\citenamefont {Lombardini}\ and\ \citenamefont
  {Poirier}(2006)}]{weylets_Ne2_poirier_2006}%
  \BibitemOpen
  \bibfield  {author} {\bibinfo {author} {\bibfnamefont {R.}~\bibnamefont
  {Lombardini}}\ and\ \bibinfo {author} {\bibfnamefont {B.}~\bibnamefont
  {Poirier}},\ }\href {\doibase 10.1063/1.2187473} {\bibfield  {journal}
  {\bibinfo  {journal} {J. Chem. Phys.}\ }\textbf {\bibinfo {volume} {124}},\
  \bibinfo {pages} {144107} (\bibinfo {year} {2006})}\BibitemShut {NoStop}%
\bibitem [{\citenamefont {Shimshovitz}(2015)}]{asaf_thesis}%
  \BibitemOpen
  \bibfield  {author} {\bibinfo {author} {\bibfnamefont {A.}~\bibnamefont
  {Shimshovitz}},\ }\emph {\bibinfo {title} {Phase Space Approach to Solving
  the Schrödinger Equation}},\ \href@noop {} {Ph.D. thesis},\ \bibinfo
  {school} {Weizmann Institute of Science} (\bibinfo {year} {2015})\BibitemShut
  {NoStop}%
\bibitem [{\citenamefont {Shimshovitz}, \citenamefont {Bačić},\ and\
  \citenamefont {Tannor}(2014)}]{pvb_LiCN_tannor_2014}%
  \BibitemOpen
  \bibfield  {author} {\bibinfo {author} {\bibfnamefont {A.}~\bibnamefont
  {Shimshovitz}}, \bibinfo {author} {\bibfnamefont {Z.}~\bibnamefont
  {Bačić}}, \ and\ \bibinfo {author} {\bibfnamefont {D.~J.}\ \bibnamefont
  {Tannor}},\ }\href {\doibase 10.1063/1.4902553} {\bibfield  {journal}
  {\bibinfo  {journal} {J. Chem. Phys.}\ }\textbf {\bibinfo {volume} {141}},\
  \bibinfo {pages} {234106} (\bibinfo {year} {2014})}\BibitemShut {NoStop}%
\bibitem [{\citenamefont {Brown}\ and\ \citenamefont
  {Carrington}(2015)}]{pvb_H2O_carrington_2015}%
  \BibitemOpen
  \bibfield  {author} {\bibinfo {author} {\bibfnamefont {J.}~\bibnamefont
  {Brown}}\ and\ \bibinfo {author} {\bibfnamefont {T.}~\bibnamefont
  {Carrington}},\ }\href {\doibase 10.1063/1.4926805} {\bibfield  {journal}
  {\bibinfo  {journal} {J. Chem. Phys.}\ }\textbf {\bibinfo {volume} {143}},\
  \bibinfo {pages} {044104} (\bibinfo {year} {2015})}\BibitemShut {NoStop}%
\bibitem [{\citenamefont {Halverson}\ and\ \citenamefont
  {Poirier}(2014)}]{symmetr_gauss_appl_CH2NH_poirier_2014}%
  \BibitemOpen
  \bibfield  {author} {\bibinfo {author} {\bibfnamefont {T.}~\bibnamefont
  {Halverson}}\ and\ \bibinfo {author} {\bibfnamefont {B.}~\bibnamefont
  {Poirier}},\ }\href {\doibase 10.1063/1.4879216} {\bibfield  {journal}
  {\bibinfo  {journal} {J. Chem. Phys.}\ }\textbf {\bibinfo {volume} {140}},\
  \bibinfo {pages} {204112} (\bibinfo {year} {2014})}\BibitemShut {NoStop}%
\bibitem [{\citenamefont {Halverson}\ and\ \citenamefont
  {Poirier}(2015{\natexlab{a}})}]{symmetrized_gaussians_acetonitrile_poirier_2015}%
  \BibitemOpen
  \bibfield  {author} {\bibinfo {author} {\bibfnamefont {T.}~\bibnamefont
  {Halverson}}\ and\ \bibinfo {author} {\bibfnamefont {B.}~\bibnamefont
  {Poirier}},\ }\href {\doibase 10.1016/j.cplett.2015.02.004} {\bibfield
  {journal} {\bibinfo  {journal} {Chem. Phys. Lett.}\ }\textbf {\bibinfo
  {volume} {624}},\ \bibinfo {pages} {37 } (\bibinfo {year}
  {2015}{\natexlab{a}})}\BibitemShut {NoStop}%
\bibitem [{\citenamefont {Halverson}\ and\ \citenamefont
  {Poirier}(2015{\natexlab{b}})}]{benzene_hybrid_truncation_scheme_HO_basis_poirier_2015}%
  \BibitemOpen
  \bibfield  {author} {\bibinfo {author} {\bibfnamefont {T.}~\bibnamefont
  {Halverson}}\ and\ \bibinfo {author} {\bibfnamefont {B.}~\bibnamefont
  {Poirier}},\ }\href {\doibase 10.1021/acs.jpca.5b07868} {\bibfield  {journal}
  {\bibinfo  {journal} {J. Phys. Chem. A}\ }\textbf {\bibinfo {volume} {119}},\
  \bibinfo {pages} {12417–12433} (\bibinfo {year}
  {2015}{\natexlab{b}})}\BibitemShut {NoStop}%
\bibitem [{\citenamefont {Brown}\ and\ \citenamefont
  {Carrington~Jr.}(2016)}]{symmetrized_gaussians_sop_carrington_2016}%
  \BibitemOpen
  \bibfield  {author} {\bibinfo {author} {\bibfnamefont {J.}~\bibnamefont
  {Brown}}\ and\ \bibinfo {author} {\bibfnamefont {T.}~\bibnamefont
  {Carrington~Jr.}},\ }\href {\doibase 10.1063/1.4954721} {\bibfield  {journal}
  {\bibinfo  {journal} {J. Chem. Phys.}\ }\textbf {\bibinfo {volume} {144}},\
  \bibinfo {pages} {244115} (\bibinfo {year} {2016})}\BibitemShut {NoStop}%
\bibitem [{\citenamefont {Burghardt}, \citenamefont {Meyer},\ and\
  \citenamefont {Cederbaum}(1999)}]{G-MCTDH_Burghardt_1999}%
  \BibitemOpen
  \bibfield  {author} {\bibinfo {author} {\bibfnamefont {I.}~\bibnamefont
  {Burghardt}}, \bibinfo {author} {\bibfnamefont {H.-D.}\ \bibnamefont
  {Meyer}}, \ and\ \bibinfo {author} {\bibfnamefont {L.~S.}\ \bibnamefont
  {Cederbaum}},\ }\href {\doibase 10.1063/1.479574} {\bibfield  {journal}
  {\bibinfo  {journal} {J. Chem. Phys.}\ }\textbf {\bibinfo {volume} {111}},\
  \bibinfo {pages} {2927} (\bibinfo {year} {1999})}\BibitemShut {NoStop}%
\bibitem [{\citenamefont {Römer}, \citenamefont {Ruckenbauer},\ and\
  \citenamefont {Burghardt}(2013)}]{G-MCTDH_layer_Burghardt_2013}%
  \BibitemOpen
  \bibfield  {author} {\bibinfo {author} {\bibfnamefont {S.}~\bibnamefont
  {Römer}}, \bibinfo {author} {\bibfnamefont {M.}~\bibnamefont {Ruckenbauer}},
  \ and\ \bibinfo {author} {\bibfnamefont {I.}~\bibnamefont {Burghardt}},\
  }\href {\doibase 10.1063/1.4788830} {\bibfield  {journal} {\bibinfo
  {journal} {J. Chem. Phys.}\ }\textbf {\bibinfo {volume} {138}},\ \bibinfo
  {pages} {064106} (\bibinfo {year} {2013})}\BibitemShut {NoStop}%
\bibitem [{\citenamefont
  {Worth}(2000)}]{mctdh_selected_configurations_worth_2000}%
  \BibitemOpen
  \bibfield  {author} {\bibinfo {author} {\bibfnamefont {G.~A.}\ \bibnamefont
  {Worth}},\ }\href {\doibase 10.1063/1.481438} {\bibfield  {journal} {\bibinfo
   {journal} {J. Chem. Phys.}\ }\textbf {\bibinfo {volume} {112}},\ \bibinfo
  {pages} {8322} (\bibinfo {year} {2000})}\BibitemShut {NoStop}%
\bibitem [{\citenamefont {Makri}\ and\ \citenamefont
  {Miller}(1987)}]{mctdscf_miller_1987}%
  \BibitemOpen
  \bibfield  {author} {\bibinfo {author} {\bibfnamefont {N.}~\bibnamefont
  {Makri}}\ and\ \bibinfo {author} {\bibfnamefont {W.~H.}\ \bibnamefont
  {Miller}},\ }\href {\doibase 10.1063/1.453501} {\bibfield  {journal}
  {\bibinfo  {journal} {J. Chem. Phys.}\ }\textbf {\bibinfo {volume} {87}},\
  \bibinfo {pages} {5781} (\bibinfo {year} {1987})}\BibitemShut {NoStop}%
\bibitem [{\citenamefont {Hammerich}, \citenamefont {Kosloff},\ and\
  \citenamefont {Ratner}(1987)}]{mctdscf_kosloff_1987}%
  \BibitemOpen
  \bibfield  {author} {\bibinfo {author} {\bibfnamefont {A.~D.}\ \bibnamefont
  {Hammerich}}, \bibinfo {author} {\bibfnamefont {R.}~\bibnamefont {Kosloff}},
  \ and\ \bibinfo {author} {\bibfnamefont {M.~A.}\ \bibnamefont {Ratner}},\
  }\enquote {\bibinfo {title} {Time dependent quantum mechanical calculations
  of the dissociation dynamics of the cluster $\rm{He}_n$-$\rm{I}_2$},}\ in\
  \href@noop {} {\emph {\bibinfo {booktitle} {Large Finite Systems: Proceedings
  of the Twentieth Jerusalem Symposium on Quantum Chemistry and Biochemistry
  Held in Jerusalem}}},\ Vol.~\bibinfo {volume} {20},\ \bibinfo {editor}
  {edited by\ \bibinfo {editor} {\bibfnamefont {J.}~\bibnamefont {Jortner}}\
  and\ \bibinfo {editor} {\bibfnamefont {A.}~\bibnamefont {Pullman}}}\
  (\bibinfo  {publisher} {Springer},\ \bibinfo {year} {1987})\BibitemShut
  {NoStop}%
\bibitem [{\citenamefont {Hammerich}, \citenamefont {Kosloff},\ and\
  \citenamefont {Ratner}(1990)}]{mctdscf_kosloff_1990}%
  \BibitemOpen
  \bibfield  {author} {\bibinfo {author} {\bibfnamefont {A.~D.}\ \bibnamefont
  {Hammerich}}, \bibinfo {author} {\bibfnamefont {R.}~\bibnamefont {Kosloff}},
  \ and\ \bibinfo {author} {\bibfnamefont {M.~A.}\ \bibnamefont {Ratner}},\
  }\href {\doibase 10.1016/0009-2614(90)80057-K} {\bibfield  {journal}
  {\bibinfo  {journal} {Chem. Phys. Lett.}\ }\textbf {\bibinfo {volume}
  {171}},\ \bibinfo {pages} {97 } (\bibinfo {year} {1990})}\BibitemShut
  {NoStop}%
\bibitem [{\citenamefont {Haxton}\ and\ \citenamefont
  {McCurdy}(2015)}]{restricted_mctdh_variational_principle_mccurdy_2015}%
  \BibitemOpen
  \bibfield  {author} {\bibinfo {author} {\bibfnamefont {D.~J.}\ \bibnamefont
  {Haxton}}\ and\ \bibinfo {author} {\bibfnamefont {C.~W.}\ \bibnamefont
  {McCurdy}},\ }\href {\doibase 10.1103/PhysRevA.91.012509} {\bibfield
  {journal} {\bibinfo  {journal} {Phys. Rev. A}\ }\textbf {\bibinfo {volume}
  {91}},\ \bibinfo {pages} {012509} (\bibinfo {year} {2015})}\BibitemShut
  {NoStop}%
\bibitem [{\citenamefont {Miyagi}\ and\ \citenamefont
  {Madsen}(2013)}]{tdRASSCF_madsen_2013}%
  \BibitemOpen
  \bibfield  {author} {\bibinfo {author} {\bibfnamefont {H.}~\bibnamefont
  {Miyagi}}\ and\ \bibinfo {author} {\bibfnamefont {L.~B.}\ \bibnamefont
  {Madsen}},\ }\href {\doibase 10.1103/PhysRevA.87.062511} {\bibfield
  {journal} {\bibinfo  {journal} {Phys. Rev. A}\ }\textbf {\bibinfo {volume}
  {87}},\ \bibinfo {pages} {062511} (\bibinfo {year} {2013})}\BibitemShut
  {NoStop}%
\bibitem [{\citenamefont {Miyagi}\ and\ \citenamefont
  {Madsen}(2014)}]{tdRASSCF_extension_madsen_2014}%
  \BibitemOpen
  \bibfield  {author} {\bibinfo {author} {\bibfnamefont {H.}~\bibnamefont
  {Miyagi}}\ and\ \bibinfo {author} {\bibfnamefont {L.~B.}\ \bibnamefont
  {Madsen}},\ }\href {\doibase 10.1103/PhysRevA.89.063416} {\bibfield
  {journal} {\bibinfo  {journal} {Phys. Rev. A}\ }\textbf {\bibinfo {volume}
  {89}},\ \bibinfo {pages} {063416} (\bibinfo {year} {2014})}\BibitemShut
  {NoStop}%
\bibitem [{\citenamefont {Miyagi}\ and\ \citenamefont
  {Madsen}(2017)}]{tdRASSCF_space_partition_madsen_2017}%
  \BibitemOpen
  \bibfield  {author} {\bibinfo {author} {\bibfnamefont {H.}~\bibnamefont
  {Miyagi}}\ and\ \bibinfo {author} {\bibfnamefont {L.~B.}\ \bibnamefont
  {Madsen}},\ }\href {\doibase 10.1103/PhysRevA.95.023415} {\bibfield
  {journal} {\bibinfo  {journal} {Phys. Rev. A}\ }\textbf {\bibinfo {volume}
  {95}},\ \bibinfo {pages} {023415} (\bibinfo {year} {2017})}\BibitemShut
  {NoStop}%
\bibitem [{\citenamefont {Sato}\ and\ \citenamefont
  {Ishikawa}(2015)}]{tdORMAS_ishikawa_2015}%
  \BibitemOpen
  \bibfield  {author} {\bibinfo {author} {\bibfnamefont {T.}~\bibnamefont
  {Sato}}\ and\ \bibinfo {author} {\bibfnamefont {K.~L.}\ \bibnamefont
  {Ishikawa}},\ }\href {\doibase 10.1103/PhysRevA.91.023417} {\bibfield
  {journal} {\bibinfo  {journal} {Phys. Rev. A}\ }\textbf {\bibinfo {volume}
  {91}},\ \bibinfo {pages} {023417} (\bibinfo {year} {2015})}\BibitemShut
  {NoStop}%
\bibitem [{\citenamefont {Lévêque}\ and\ \citenamefont
  {Madsen}(2017)}]{td_rasscf_bosons_madsen_2017}%
  \BibitemOpen
  \bibfield  {author} {\bibinfo {author} {\bibfnamefont {C.}~\bibnamefont
  {Lévêque}}\ and\ \bibinfo {author} {\bibfnamefont {L.~B.}\ \bibnamefont
  {Madsen}},\ }\href {\doibase 10.1088/1367-2630/aa6319} {\bibfield  {journal}
  {\bibinfo  {journal} {New J. Phy.}\ }\textbf {\bibinfo {volume} {19}},\
  \bibinfo {pages} {043007} (\bibinfo {year} {2017})}\BibitemShut {NoStop}%
\bibitem [{\citenamefont {Worth}, \citenamefont {Robb},\ and\ \citenamefont
  {Burghardt}(2004)}]{vMCG_Burghardt_2004}%
  \BibitemOpen
  \bibfield  {author} {\bibinfo {author} {\bibfnamefont {G.~A.}\ \bibnamefont
  {Worth}}, \bibinfo {author} {\bibfnamefont {M.~A.}\ \bibnamefont {Robb}}, \
  and\ \bibinfo {author} {\bibfnamefont {I.}~\bibnamefont {Burghardt}},\ }\href
  {\doibase 10.1039/b314253a} {\bibfield  {journal} {\bibinfo  {journal}
  {Faraday Discuss.}\ }\textbf {\bibinfo {volume} {127}},\ \bibinfo {pages}
  {307} (\bibinfo {year} {2004})}\BibitemShut {NoStop}%
\bibitem [{\citenamefont {Richings}\ \emph {et~al.}(2015)\citenamefont
  {Richings}, \citenamefont {Polyak}, \citenamefont {Spinlove}, \citenamefont
  {Worth}, \citenamefont {Burghardt},\ and\ \citenamefont
  {Lasorne}}]{vMCG_rev_lasorne_2015}%
  \BibitemOpen
  \bibfield  {author} {\bibinfo {author} {\bibfnamefont {G.}~\bibnamefont
  {Richings}}, \bibinfo {author} {\bibfnamefont {I.}~\bibnamefont {Polyak}},
  \bibinfo {author} {\bibfnamefont {K.}~\bibnamefont {Spinlove}}, \bibinfo
  {author} {\bibfnamefont {G.}~\bibnamefont {Worth}}, \bibinfo {author}
  {\bibfnamefont {I.}~\bibnamefont {Burghardt}}, \ and\ \bibinfo {author}
  {\bibfnamefont {B.}~\bibnamefont {Lasorne}},\ }\href {\doibase
  10.1080/0144235x.2015.1051354} {\bibfield  {journal} {\bibinfo  {journal}
  {Int. Rev. Phys. Chem.}\ }\textbf {\bibinfo {volume} {34}},\ \bibinfo {pages}
  {269–308} (\bibinfo {year} {2015})}\BibitemShut {NoStop}%
\bibitem [{\citenamefont {Meyer}\ \emph {et~al.}(2006)\citenamefont {Meyer},
  \citenamefont {Quéré}, \citenamefont {Léonard},\ and\ \citenamefont
  {Gatti}}]{mctdh_improved_diagonalisation_gatti_2016}%
  \BibitemOpen
  \bibfield  {author} {\bibinfo {author} {\bibfnamefont {H.-D.}\ \bibnamefont
  {Meyer}}, \bibinfo {author} {\bibfnamefont {F.~L.}\ \bibnamefont {Quéré}},
  \bibinfo {author} {\bibfnamefont {C.}~\bibnamefont {Léonard}}, \ and\
  \bibinfo {author} {\bibfnamefont {F.}~\bibnamefont {Gatti}},\ }\href
  {\doibase 10.1016/j.chemphys.2006.06.002} {\bibfield  {journal} {\bibinfo
  {journal} {Chem. Phys.}\ }\textbf {\bibinfo {volume} {329}},\ \bibinfo
  {pages} {179 } (\bibinfo {year} {2006})}\BibitemShut {NoStop}%
\bibitem [{\citenamefont {Culot}, \citenamefont {Laruelle},\ and\ \citenamefont
  {Li{\'e}vin}(1995)}]{vCASSCF_lievin_1995}%
  \BibitemOpen
  \bibfield  {author} {\bibinfo {author} {\bibfnamefont {F.}~\bibnamefont
  {Culot}}, \bibinfo {author} {\bibfnamefont {F.}~\bibnamefont {Laruelle}}, \
  and\ \bibinfo {author} {\bibfnamefont {J.}~\bibnamefont {Li{\'e}vin}},\
  }\href {\doibase 10.1007/BF01125947} {\bibfield  {journal} {\bibinfo
  {journal} {Theor. Chim. Acta}\ }\textbf {\bibinfo {volume} {92}},\ \bibinfo
  {pages} {211} (\bibinfo {year} {1995})}\BibitemShut {NoStop}%
\bibitem [{\citenamefont {Culot}\ and\ \citenamefont
  {Li{\'e}vin}(1994)}]{vMCSCF_lievin_1994}%
  \BibitemOpen
  \bibfield  {author} {\bibinfo {author} {\bibfnamefont {F.}~\bibnamefont
  {Culot}}\ and\ \bibinfo {author} {\bibfnamefont {J.}~\bibnamefont
  {Li{\'e}vin}},\ }\href {\doibase 10.1007/BF01225116} {\bibfield  {journal}
  {\bibinfo  {journal} {Theor. Chim. Acta}\ }\textbf {\bibinfo {volume} {89}},\
  \bibinfo {pages} {227} (\bibinfo {year} {1994})}\BibitemShut {NoStop}%
\bibitem [{\citenamefont {Heislbetz}\ and\ \citenamefont
  {Rauhut}(2010)}]{vMCSCF_rauhut_2010}%
  \BibitemOpen
  \bibfield  {author} {\bibinfo {author} {\bibfnamefont {S.}~\bibnamefont
  {Heislbetz}}\ and\ \bibinfo {author} {\bibfnamefont {G.}~\bibnamefont
  {Rauhut}},\ }\href {\doibase 10.1063/1.3364861} {\bibfield  {journal}
  {\bibinfo  {journal} {J. Chem. Phys.}\ }\textbf {\bibinfo {volume} {132}},\
  \bibinfo {pages} {124102} (\bibinfo {year} {2010})}\BibitemShut {NoStop}%
\bibitem [{\citenamefont {Heislbetz}, \citenamefont {Pfeiffer},\ and\
  \citenamefont {Rauhut}(2011)}]{vMCSCF_pruned_rauhut_2010}%
  \BibitemOpen
  \bibfield  {author} {\bibinfo {author} {\bibfnamefont {S.}~\bibnamefont
  {Heislbetz}}, \bibinfo {author} {\bibfnamefont {F.}~\bibnamefont {Pfeiffer}},
  \ and\ \bibinfo {author} {\bibfnamefont {G.}~\bibnamefont {Rauhut}},\ }\href
  {\doibase 10.1063/1.3593714} {\bibfield  {journal} {\bibinfo  {journal} {J.
  Chem. Phys.}\ }\textbf {\bibinfo {volume} {134}},\ \bibinfo {pages} {204108}
  (\bibinfo {year} {2011})}\BibitemShut {NoStop}%
\bibitem [{\citenamefont {Mizukami}\ and\ \citenamefont
  {Tew}(2013)}]{vMRPT2_pruned_tew_2013}%
  \BibitemOpen
  \bibfield  {author} {\bibinfo {author} {\bibfnamefont {W.}~\bibnamefont
  {Mizukami}}\ and\ \bibinfo {author} {\bibfnamefont {D.~P.}\ \bibnamefont
  {Tew}},\ }\href {\doibase 10.1063/1.4830100} {\bibfield  {journal} {\bibinfo
  {journal} {J. Chem. Phys.}\ }\textbf {\bibinfo {volume} {139}},\ \bibinfo
  {pages} {194108} (\bibinfo {year} {2013})}\BibitemShut {NoStop}%
\bibitem [{\citenamefont {Meier}\ \emph {et~al.}(2015)\citenamefont {Meier},
  \citenamefont {Oschetzki}, \citenamefont {Pfeiffer},\ and\ \citenamefont
  {Rauhut}}]{vMCSCF_pruned_state_averaged_rauhut_2015}%
  \BibitemOpen
  \bibfield  {author} {\bibinfo {author} {\bibfnamefont {P.}~\bibnamefont
  {Meier}}, \bibinfo {author} {\bibfnamefont {D.}~\bibnamefont {Oschetzki}},
  \bibinfo {author} {\bibfnamefont {F.}~\bibnamefont {Pfeiffer}}, \ and\
  \bibinfo {author} {\bibfnamefont {G.}~\bibnamefont {Rauhut}},\ }\href
  {\doibase 10.1063/1.4938280} {\bibfield  {journal} {\bibinfo  {journal} {J.
  Chem. Phys.}\ }\textbf {\bibinfo {volume} {143}},\ \bibinfo {pages} {244111}
  (\bibinfo {year} {2015})}\BibitemShut {NoStop}%
\bibitem [{\citenamefont {Pfeiffer}\ and\ \citenamefont
  {Rauhut}(2014)}]{vMCSCF_pruned_perturbation_theory_rauhut_2014}%
  \BibitemOpen
  \bibfield  {author} {\bibinfo {author} {\bibfnamefont {F.}~\bibnamefont
  {Pfeiffer}}\ and\ \bibinfo {author} {\bibfnamefont {G.}~\bibnamefont
  {Rauhut}},\ }\href {\doibase 10.1063/1.4865098} {\bibfield  {journal}
  {\bibinfo  {journal} {J. Chem. Phys.}\ }\textbf {\bibinfo {volume} {140}},\
  \bibinfo {pages} {064110} (\bibinfo {year} {2014})}\BibitemShut {NoStop}%
\bibitem [{\citenamefont {Wodraszka}\ and\ \citenamefont
  {Carrington}(2016)}]{pruned_mctdh_carrington_2016}%
  \BibitemOpen
  \bibfield  {author} {\bibinfo {author} {\bibfnamefont {R.}~\bibnamefont
  {Wodraszka}}\ and\ \bibinfo {author} {\bibfnamefont {T.}~\bibnamefont
  {Carrington}},\ }\href {\doibase 10.1063/1.4959228} {\bibfield  {journal}
  {\bibinfo  {journal} {J. Chem. Phys.}\ }\textbf {\bibinfo {volume} {145}},\
  \bibinfo {pages} {044110} (\bibinfo {year} {2016})}\BibitemShut {NoStop}%
\bibitem [{\citenamefont {Wodraszka}\ and\ \citenamefont
  {Carrington}(2017)}]{wodraszka_2017}%
  \BibitemOpen
  \bibfield  {author} {\bibinfo {author} {\bibfnamefont {R.}~\bibnamefont
  {Wodraszka}}\ and\ \bibinfo {author} {\bibfnamefont {T.}~\bibnamefont
  {Carrington}},\ }\href {\doibase 10.1063/1.4983281} {\bibfield  {journal}
  {\bibinfo  {journal} {J. Chem. Phys.}\ }\textbf {\bibinfo {volume} {146}},\
  \bibinfo {pages} {194105} (\bibinfo {year} {2017})}\BibitemShut {NoStop}%
\bibitem [{\citenamefont {Machnes}, \citenamefont {Assémat},\ and\
  \citenamefont {Tannor}(2016)}]{pvb_algorithms_tannor_2016}%
  \BibitemOpen
  \bibfield  {author} {\bibinfo {author} {\bibfnamefont {S.}~\bibnamefont
  {Machnes}}, \bibinfo {author} {\bibfnamefont {E.}~\bibnamefont {Assémat}}, \
  and\ \bibinfo {author} {\bibfnamefont {D.}~\bibnamefont {Tannor}},\
  }\href@noop {} {\enquote {\bibinfo {title} {{Quantum Dynamics in Phase space
  using the Biorthogonal von Neumann bases: Algorithmic Considerations}},}\ }
  (\bibinfo {year} {2016}),\ \Eprint {http://arxiv.org/abs/arXiv:1603.03963}
  {arXiv:1603.03963} \BibitemShut {NoStop}%
\bibitem [{\citenamefont {Kolda}\ and\ \citenamefont
  {Bader}(2009)}]{tensor_decomp_rev_bader_2009}%
  \BibitemOpen
  \bibfield  {author} {\bibinfo {author} {\bibfnamefont {T.~G.}\ \bibnamefont
  {Kolda}}\ and\ \bibinfo {author} {\bibfnamefont {B.~W.}\ \bibnamefont
  {Bader}},\ }\href {\doibase 10.1137/07070111X} {\bibfield  {journal}
  {\bibinfo  {journal} {SIAM Rev.}\ }\textbf {\bibinfo {volume} {51}},\
  \bibinfo {pages} {455} (\bibinfo {year} {2009})}\BibitemShut {NoStop}%
\bibitem [{\citenamefont {Helgaker}, \citenamefont {Olsen},\ and\ \citenamefont
  {Jørgensen}(2013)}]{helgaker_book}%
  \BibitemOpen
  \bibfield  {author} {\bibinfo {author} {\bibfnamefont {T.}~\bibnamefont
  {Helgaker}}, \bibinfo {author} {\bibfnamefont {J.}~\bibnamefont {Olsen}}, \
  and\ \bibinfo {author} {\bibfnamefont {P.}~\bibnamefont {Jørgensen}},\
  }\href@noop {} {\emph {\bibinfo {title} {Molecular Electronic-Structure
  Theory}}},\ \bibinfo {edition} {1st}\ ed.\ (\bibinfo  {publisher} {Wiley},\
  \bibinfo {year} {2013})\BibitemShut {NoStop}%
\bibitem [{\citenamefont {Dirac}(1930)}]{td_variational_principle_dirac_1930}%
  \BibitemOpen
  \bibfield  {author} {\bibinfo {author} {\bibfnamefont {P.~A.~M.}\
  \bibnamefont {Dirac}},\ }\href {\doibase 10.1017/S0305004100016108}
  {\bibfield  {journal} {\bibinfo  {journal} {Proc. Camb. Phil. Soc.}\ }\textbf
  {\bibinfo {volume} {26}},\ \bibinfo {pages} {376–385} (\bibinfo {year}
  {1930})}\BibitemShut {NoStop}%
\bibitem [{\citenamefont
  {Frenkel}(1934)}]{td_variational_principle_frenkel_1934}%
  \BibitemOpen
  \bibfield  {author} {\bibinfo {author} {\bibfnamefont {J.}~\bibnamefont
  {Frenkel}},\ }\href@noop {} {\emph {\bibinfo {title} {Wave Mechanics --
  Advanced General Theory}}}\ (\bibinfo  {publisher} {Oxford at the Clarendon
  Press},\ \bibinfo {year} {1934})\BibitemShut {NoStop}%
\bibitem [{\citenamefont
  {McLachlan}(1964)}]{td_variational_principle_mclachlan_1964}%
  \BibitemOpen
  \bibfield  {author} {\bibinfo {author} {\bibfnamefont {A.}~\bibnamefont
  {McLachlan}},\ }\href {\doibase 10.1080/00268976400100041} {\bibfield
  {journal} {\bibinfo  {journal} {Mol. Phys.}\ }\textbf {\bibinfo {volume}
  {8}},\ \bibinfo {pages} {39} (\bibinfo {year} {1964})}\BibitemShut {NoStop}%
\bibitem [{\citenamefont {Broeckhove}\ \emph {et~al.}(1988)\citenamefont
  {Broeckhove}, \citenamefont {Lathouwers}, \citenamefont {Kesteloot},\ and\
  \citenamefont {Leuven}}]{td_var_principle_equivalence_van-leuven_1988}%
  \BibitemOpen
  \bibfield  {author} {\bibinfo {author} {\bibfnamefont {J.}~\bibnamefont
  {Broeckhove}}, \bibinfo {author} {\bibfnamefont {L.}~\bibnamefont
  {Lathouwers}}, \bibinfo {author} {\bibfnamefont {E.}~\bibnamefont
  {Kesteloot}}, \ and\ \bibinfo {author} {\bibfnamefont {P.~V.}\ \bibnamefont
  {Leuven}},\ }\href {\doibase 10.1016/0009-2614(88)80380-4} {\bibfield
  {journal} {\bibinfo  {journal} {Chem. Phys. Lett.}\ }\textbf {\bibinfo
  {volume} {149}},\ \bibinfo {pages} {547 } (\bibinfo {year}
  {1988})}\BibitemShut {NoStop}%
\bibitem [{\citenamefont {Hinz}, \citenamefont {Bauch},\ and\ \citenamefont
  {Bonitz}(2016)}]{mctdh_denmat_inv_inacc_bonitz_2016}%
  \BibitemOpen
  \bibfield  {author} {\bibinfo {author} {\bibfnamefont {C.~M.}\ \bibnamefont
  {Hinz}}, \bibinfo {author} {\bibfnamefont {S.}~\bibnamefont {Bauch}}, \ and\
  \bibinfo {author} {\bibfnamefont {M.}~\bibnamefont {Bonitz}},\ }\href
  {http://stacks.iop.org/1742-6596/696/i=1/a=012009} {\bibfield  {journal}
  {\bibinfo  {journal} {J. Phys. Conf. Ser.}\ }\textbf {\bibinfo {volume}
  {696}},\ \bibinfo {pages} {012009} (\bibinfo {year} {2016})}\BibitemShut
  {NoStop}%
\bibitem [{\citenamefont {Lubich}(2015)}]{mctdh_denmat_inv_lubich_2015}%
  \BibitemOpen
  \bibfield  {author} {\bibinfo {author} {\bibfnamefont {C.}~\bibnamefont
  {Lubich}},\ }\href {\doibase 10.1093/amrx/abv006} {\bibfield  {journal}
  {\bibinfo  {journal} {Appl. Math. Res. Express}\ ,\ \bibinfo {pages} {311}}
  (\bibinfo {year} {2015})}\BibitemShut {NoStop}%
\bibitem [{\citenamefont {Kloss}, \citenamefont {Burghardt},\ and\
  \citenamefont {Lubich}(2017)}]{mctdh_denmat_inv_lubich_2017}%
  \BibitemOpen
  \bibfield  {author} {\bibinfo {author} {\bibfnamefont {B.}~\bibnamefont
  {Kloss}}, \bibinfo {author} {\bibfnamefont {I.}~\bibnamefont {Burghardt}}, \
  and\ \bibinfo {author} {\bibfnamefont {C.}~\bibnamefont {Lubich}},\ }\href
  {\doibase 10.1063/1.4982065} {\bibfield  {journal} {\bibinfo  {journal} {J.
  Chem. Phys.}\ }\textbf {\bibinfo {volume} {146}},\ \bibinfo {pages} {174107}
  (\bibinfo {year} {2017})}\BibitemShut {NoStop}%
\bibitem [{\citenamefont {Lee}\ and\ \citenamefont
  {Fischer}(2014)}]{mctdh_denmat_inv_fischer_2014}%
  \BibitemOpen
  \bibfield  {author} {\bibinfo {author} {\bibfnamefont {K.-S.}\ \bibnamefont
  {Lee}}\ and\ \bibinfo {author} {\bibfnamefont {U.~R.}\ \bibnamefont
  {Fischer}},\ }\href {\doibase 10.1142/S0217979215500216} {\bibfield
  {journal} {\bibinfo  {journal} {Int. J. Mod. Phys. B}\ }\textbf {\bibinfo
  {volume} {28}},\ \bibinfo {pages} {1550021} (\bibinfo {year}
  {2014})}\BibitemShut {NoStop}%
\bibitem [{\citenamefont {Baye}\ and\ \citenamefont
  {Heenen}(1986)}]{dvr_origin_heenen_1986}%
  \BibitemOpen
  \bibfield  {author} {\bibinfo {author} {\bibfnamefont {D.}~\bibnamefont
  {Baye}}\ and\ \bibinfo {author} {\bibfnamefont {P.~H.}\ \bibnamefont
  {Heenen}},\ }\href {http://stacks.iop.org/0305-4470/19/i=11/a=013} {\bibfield
   {journal} {\bibinfo  {journal} {J. Phys. A: Math. Gen.}\ }\textbf {\bibinfo
  {volume} {19}},\ \bibinfo {pages} {2041} (\bibinfo {year}
  {1986})}\BibitemShut {NoStop}%
\bibitem [{\citenamefont {Baye}(2015)}]{lagrange_mesh_rev_baye_2015}%
  \BibitemOpen
  \bibfield  {author} {\bibinfo {author} {\bibfnamefont {D.}~\bibnamefont
  {Baye}},\ }\href {\doibase 10.1016/j.physrep.2014.11.006} {\bibfield
  {journal} {\bibinfo  {journal} {Phys. Rep.}\ }\textbf {\bibinfo {volume}
  {565}},\ \bibinfo {pages} {1 } (\bibinfo {year} {2015})}\BibitemShut
  {NoStop}%
\bibitem [{Note1()}]{Note1}%
  \BibitemOpen
  \bibinfo {note} {For mode combination, the scaling in computational effort
  for the description of the SPFs can be implemented as $g D n^2 M^{P+1}$,
  where $P$ is the dimension of the SPFs and $M$ is the geometric mean of the
  numbers of one-dimensional primitive functions describing the SPFs. This is
  often more favorable than a $g D n M^{2P} = g D n N^2$ scaling.}\BibitemShut
  {Stop}%
\bibitem [{\citenamefont {Hackbusch}\ and\ \citenamefont
  {K{\"u}hn}(2009)}]{hierarchical_tucker_decomp_kuehn_2009}%
  \BibitemOpen
  \bibfield  {author} {\bibinfo {author} {\bibfnamefont {W.}~\bibnamefont
  {Hackbusch}}\ and\ \bibinfo {author} {\bibfnamefont {S.}~\bibnamefont
  {K{\"u}hn}},\ }\href {\doibase 10.1007/s00041-009-9094-9} {\bibfield
  {journal} {\bibinfo  {journal} {J. Fourier Anal. Appl.}\ }\textbf {\bibinfo
  {volume} {15}},\ \bibinfo {pages} {706} (\bibinfo {year} {2009})}\BibitemShut
  {NoStop}%
\bibitem [{\citenamefont
  {Grasedyck}(2010)}]{hierarchical_tucker_decomp_grasedyck_2010}%
  \BibitemOpen
  \bibfield  {author} {\bibinfo {author} {\bibfnamefont {L.}~\bibnamefont
  {Grasedyck}},\ }\href {\doibase 10.1137/090764189} {\bibfield  {journal}
  {\bibinfo  {journal} {SIAM. J. Matrix Anal. and Appl.}\ }\textbf {\bibinfo
  {volume} {31}},\ \bibinfo {pages} {2029} (\bibinfo {year}
  {2010})}\BibitemShut {NoStop}%
\bibitem [{\citenamefont {Sielk}\ \emph {et~al.}(2009)\citenamefont {Sielk},
  \citenamefont {von Horsten}, \citenamefont {Krüger}, \citenamefont
  {Schneider},\ and\ \citenamefont {Hartke}}]{proDG_hartke_2008}%
  \BibitemOpen
  \bibfield  {author} {\bibinfo {author} {\bibfnamefont {J.}~\bibnamefont
  {Sielk}}, \bibinfo {author} {\bibfnamefont {H.~F.}\ \bibnamefont {von
  Horsten}}, \bibinfo {author} {\bibfnamefont {F.}~\bibnamefont {Krüger}},
  \bibinfo {author} {\bibfnamefont {R.}~\bibnamefont {Schneider}}, \ and\
  \bibinfo {author} {\bibfnamefont {B.}~\bibnamefont {Hartke}},\ }\href
  {\doibase 10.1039/b814315c} {\bibfield  {journal} {\bibinfo  {journal} {Phys.
  Chem. Chem. Phys.}\ }\textbf {\bibinfo {volume} {11}},\ \bibinfo {pages}
  {463–475} (\bibinfo {year} {2009})}\BibitemShut {NoStop}%
\bibitem [{\citenamefont {Jansen}(1993)}]{mctdh_natural_orbitals_jensen_1993}%
  \BibitemOpen
  \bibfield  {author} {\bibinfo {author} {\bibfnamefont {A.~P.~J.}\
  \bibnamefont {Jansen}},\ }\href {\doibase 10.1063/1.466101} {\bibfield
  {journal} {\bibinfo  {journal} {J. Chem. Phys.}\ }\textbf {\bibinfo {volume}
  {99}},\ \bibinfo {pages} {4055} (\bibinfo {year} {1993})}\BibitemShut
  {NoStop}%
\bibitem [{\citenamefont {Manthe}(1994)}]{mctdh_natural_orbitals_manthe_1994}%
  \BibitemOpen
  \bibfield  {author} {\bibinfo {author} {\bibfnamefont {U.}~\bibnamefont
  {Manthe}},\ }\href {\doibase 10.1063/1.467644} {\bibfield  {journal}
  {\bibinfo  {journal} {J. Chem. Phys.}\ }\textbf {\bibinfo {volume} {101}},\
  \bibinfo {pages} {2652} (\bibinfo {year} {1994})}\BibitemShut {NoStop}%
\bibitem [{\citenamefont {L\"owdin}(1955)}]{natural_orbitals_lowdin_1955}%
  \BibitemOpen
  \bibfield  {author} {\bibinfo {author} {\bibfnamefont {P.-O.}\ \bibnamefont
  {L\"owdin}},\ }\href {\doibase 10.1103/PhysRev.97.1474} {\bibfield  {journal}
  {\bibinfo  {journal} {Phys. Rev.}\ }\textbf {\bibinfo {volume} {97}},\
  \bibinfo {pages} {1474} (\bibinfo {year} {1955})}\BibitemShut {NoStop}%
\bibitem [{\citenamefont {Guennebaud}, \citenamefont {Jacob}\ \emph
  {et~al.}(2017)\citenamefont {Guennebaud}, \citenamefont {Jacob} \emph
  {et~al.}}]{eigen_lib}%
  \BibitemOpen
  \bibfield  {author} {\bibinfo {author} {\bibfnamefont {G.}~\bibnamefont
  {Guennebaud}}, \bibinfo {author} {\bibfnamefont {B.}~\bibnamefont {Jacob}},
  \emph {et~al.},\ }\href@noop {} {\enquote {\bibinfo {title} {Eigen v3.3},}\
  }\bibinfo {howpublished} {http://eigen.tuxfamily.org} (\bibinfo {year}
  {2017})\BibitemShut {NoStop}%
\bibitem [{\citenamefont {{Intel Corporation}}(2016)}]{mkl_lib}%
  \BibitemOpen
  \bibfield  {author} {\bibinfo {author} {\bibnamefont {{Intel Corporation}}},\
  }\href@noop {} {\enquote {\bibinfo {title} {Intel® math kernel library,
  version 11.3.3},}\ }\bibinfo {howpublished}
  {http://software.intel.com/en-us/articles/intel-mkl/} (\bibinfo {year}
  {2016})\BibitemShut {NoStop}%
\bibitem [{\citenamefont {Friesner}\ \emph {et~al.}(1986)\citenamefont
  {Friesner}, \citenamefont {Wyatt}, \citenamefont {Hempel},\ and\
  \citenamefont {Criner}}]{DVR_matvec_calc_friesner_1986}%
  \BibitemOpen
  \bibfield  {author} {\bibinfo {author} {\bibfnamefont {R.~A.}\ \bibnamefont
  {Friesner}}, \bibinfo {author} {\bibfnamefont {R.~E.}\ \bibnamefont {Wyatt}},
  \bibinfo {author} {\bibfnamefont {C.}~\bibnamefont {Hempel}}, \ and\ \bibinfo
  {author} {\bibfnamefont {B.}~\bibnamefont {Criner}},\ }\href {\doibase
  10.1016/0021-9991(86)90026-4} {\bibfield  {journal} {\bibinfo  {journal} {J.
  Comp. Phys.}\ }\textbf {\bibinfo {volume} {64}},\ \bibinfo {pages} {220 }
  (\bibinfo {year} {1986})}\BibitemShut {NoStop}%
\bibitem [{\citenamefont {Manthe}\ and\ \citenamefont
  {Köppel}(1990)}]{DVR_matvec_calc_manthe_1990}%
  \BibitemOpen
  \bibfield  {author} {\bibinfo {author} {\bibfnamefont {U.}~\bibnamefont
  {Manthe}}\ and\ \bibinfo {author} {\bibfnamefont {H.}~\bibnamefont
  {Köppel}},\ }\href {\doibase 10.1063/1.459606} {\bibfield  {journal}
  {\bibinfo  {journal} {J. Chem. Phys.}\ }\textbf {\bibinfo {volume} {93}},\
  \bibinfo {pages} {345} (\bibinfo {year} {1990})}\BibitemShut {NoStop}%
\bibitem [{\citenamefont {Bramley}\ and\ \citenamefont
  {Carrington}(1993)}]{DVR_matvec_calc_carrington_1993}%
  \BibitemOpen
  \bibfield  {author} {\bibinfo {author} {\bibfnamefont {M.~J.}\ \bibnamefont
  {Bramley}}\ and\ \bibinfo {author} {\bibfnamefont {T.}~\bibnamefont
  {Carrington}},\ }\href {\doibase 10.1063/1.465576} {\bibfield  {journal}
  {\bibinfo  {journal} {J. Chem. Phys.}\ }\textbf {\bibinfo {volume} {99}},\
  \bibinfo {pages} {8519} (\bibinfo {year} {1993})}\BibitemShut {NoStop}%
\bibitem [{\citenamefont {Cooper}\ and\ \citenamefont
  {Carrington}(2009)}]{pruned_prod_basis_mapping_carrington_2009}%
  \BibitemOpen
  \bibfield  {author} {\bibinfo {author} {\bibfnamefont {J.}~\bibnamefont
  {Cooper}}\ and\ \bibinfo {author} {\bibfnamefont {T.}~\bibnamefont
  {Carrington}},\ }\href {\doibase 10.1063/1.3140272} {\bibfield  {journal}
  {\bibinfo  {journal} {J. Chem. Phys.}\ }\textbf {\bibinfo {volume} {130}},\
  \bibinfo {pages} {214110} (\bibinfo {year} {2009})}\BibitemShut {NoStop}%
\bibitem [{\citenamefont {Light}\ and\ \citenamefont
  {Carrington}(2007)}]{dvr_rev_carrington_2007}%
  \BibitemOpen
  \bibfield  {author} {\bibinfo {author} {\bibfnamefont {J.~C.}\ \bibnamefont
  {Light}}\ and\ \bibinfo {author} {\bibfnamefont {T.}~\bibnamefont
  {Carrington}},\ }\enquote {\bibinfo {title} {{Discrete-Variable
  Representations and their Utilization}},}\ in\ \href {\doibase
  10.1002/9780470141731.ch4} {\emph {\bibinfo {booktitle} {Advances in Chemical
  Physics}}}\ (\bibinfo  {publisher} {John Wiley \& Sons, Inc.},\ \bibinfo
  {year} {2007})\ pp.\ \bibinfo {pages} {263--310}\BibitemShut {NoStop}%
\bibitem [{\citenamefont
  {Löwdin}(1950)}]{lowdin_orthogonalization_lowdin_1950}%
  \BibitemOpen
  \bibfield  {author} {\bibinfo {author} {\bibfnamefont {P.}~\bibnamefont
  {Löwdin}},\ }\href {\doibase 10.1063/1.1747632} {\bibfield  {journal}
  {\bibinfo  {journal} {J. Chem. Phys.}\ }\textbf {\bibinfo {volume} {18}},\
  \bibinfo {pages} {365} (\bibinfo {year} {1950})}\BibitemShut {NoStop}%
\bibitem [{\citenamefont {Beck}\ and\ \citenamefont
  {Meyer}(1997)}]{mctdh_cmf_meyer_1997}%
  \BibitemOpen
  \bibfield  {author} {\bibinfo {author} {\bibfnamefont {M.}~\bibnamefont
  {Beck}}\ and\ \bibinfo {author} {\bibfnamefont {H.-D.}\ \bibnamefont
  {Meyer}},\ }\href {\doibase 10.1007/s004600050342} {\bibfield  {journal}
  {\bibinfo  {journal} {Z. Phys. D}\ }\textbf {\bibinfo {volume} {42}},\
  \bibinfo {pages} {113} (\bibinfo {year} {1997})}\BibitemShut {NoStop}%
\bibitem [{\citenamefont {Stoer}\ and\ \citenamefont
  {Bulirsch}(2006)}]{bulirsch-stoer_book2}%
  \BibitemOpen
  \bibfield  {author} {\bibinfo {author} {\bibfnamefont {J.}~\bibnamefont
  {Stoer}}\ and\ \bibinfo {author} {\bibfnamefont {R.}~\bibnamefont
  {Bulirsch}},\ }\href@noop {} {\emph {\bibinfo {title} {{Numerische Mathematik
  2}}}},\ \bibinfo {edition} {5th}\ ed.\ (\bibinfo  {publisher} {Springer},\
  \bibinfo {year} {2006})\BibitemShut {NoStop}%
\bibitem [{\citenamefont {Park}\ and\ \citenamefont
  {Light}(1986)}]{sil_light_1986}%
  \BibitemOpen
  \bibfield  {author} {\bibinfo {author} {\bibfnamefont {T.~J.}\ \bibnamefont
  {Park}}\ and\ \bibinfo {author} {\bibfnamefont {J.~C.}\ \bibnamefont
  {Light}},\ }\href {\doibase 10.1063/1.451548} {\bibfield  {journal} {\bibinfo
   {journal} {J. Chem. Phys.}\ }\textbf {\bibinfo {volume} {85}},\ \bibinfo
  {pages} {5870} (\bibinfo {year} {1986})}\BibitemShut {NoStop}%
\bibitem [{\citenamefont {Worth}\ \emph {et~al.}(2014)\citenamefont {Worth},
  \citenamefont {Beck}, \citenamefont {J{\"a}ckle},\ and\ \citenamefont
  {Meyer}}]{mctdh_package}%
  \BibitemOpen
  \bibfield  {author} {\bibinfo {author} {\bibfnamefont {G.~A.}\ \bibnamefont
  {Worth}}, \bibinfo {author} {\bibfnamefont {M.~H.}\ \bibnamefont {Beck}},
  \bibinfo {author} {\bibfnamefont {A.}~\bibnamefont {J{\"a}ckle}}, \ and\
  \bibinfo {author} {\bibfnamefont {H.-D.}\ \bibnamefont {Meyer}},\ }\href@noop
  {} {}\bibinfo {howpublished} {The {MCTDH} {P}ackage, {V}ersion 8.4.10. {S}ee
  http://mctdh.uni-hd.de} (\bibinfo {year} {2014})\BibitemShut {NoStop}%
\bibitem [{\citenamefont {Manthe}, \citenamefont {Meyer},\ and\ \citenamefont
  {Cederbaum}(1992{\natexlab{b}})}]{NO2_mctdh_cederbaum_1992}%
  \BibitemOpen
  \bibfield  {author} {\bibinfo {author} {\bibfnamefont {U.}~\bibnamefont
  {Manthe}}, \bibinfo {author} {\bibfnamefont {H.-D.}\ \bibnamefont {Meyer}}, \
  and\ \bibinfo {author} {\bibfnamefont {L.~S.}\ \bibnamefont {Cederbaum}},\
  }\href {\doibase 10.1063/1.463332} {\bibfield  {journal} {\bibinfo  {journal}
  {J. Chem. Phys.}\ }\textbf {\bibinfo {volume} {97}},\ \bibinfo {pages} {9062}
  (\bibinfo {year} {1992}{\natexlab{b}})}\BibitemShut {NoStop}%
\bibitem [{\citenamefont {Marston}\ and\ \citenamefont
  {Balint-Kurti}(1989)}]{fgh_marston_balint-kurti_1989}%
  \BibitemOpen
  \bibfield  {author} {\bibinfo {author} {\bibfnamefont {C.~C.}\ \bibnamefont
  {Marston}}\ and\ \bibinfo {author} {\bibfnamefont {G.~G.}\ \bibnamefont
  {Balint-Kurti}},\ }\href {\doibase 10.1063/1.456888} {\bibfield  {journal}
  {\bibinfo  {journal} {J. Chem. Phys.}\ }\textbf {\bibinfo {volume} {91}},\
  \bibinfo {pages} {3571} (\bibinfo {year} {1989})}\BibitemShut {NoStop}%
\bibitem [{\citenamefont {Meyer}(2017)}]{meyer_communication_2017}%
  \BibitemOpen
  \bibfield  {author} {\bibinfo {author} {\bibfnamefont {H.-D.}\ \bibnamefont
  {Meyer}},\ }\href@noop {} {\enquote {\bibinfo {title} {private
  communication},}\ } (\bibinfo {year} {2017})\BibitemShut {NoStop}%
\bibitem [{\citenamefont {Meyer}\ and\ \citenamefont
  {Brill}(2009)}]{mctdh_book_parallelization}%
  \BibitemOpen
  \bibfield  {author} {\bibinfo {author} {\bibfnamefont {H.-D.}\ \bibnamefont
  {Meyer}}\ and\ \bibinfo {author} {\bibfnamefont {B.}~\bibnamefont {Brill}},\
  }\enquote {\bibinfo {title} {Shared memory parallelization of the
  multiconfiguration time-dependent hartree method},}\ in\ \href@noop {} {\emph
  {\bibinfo {booktitle} {Multidimensional Quantum Dynamics}}},\ \bibinfo
  {editor} {edited by\ \bibinfo {editor} {\bibfnamefont {H.-D.}\ \bibnamefont
  {Meyer}}, \bibinfo {editor} {\bibfnamefont {F.}~\bibnamefont {Gatti}}, \ and\
  \bibinfo {editor} {\bibfnamefont {G.~A.}\ \bibnamefont {Worth}}}\ (\bibinfo
  {publisher} {Wiley-VCH},\ \bibinfo {year} {2009})\ \bibinfo {edition} {1st}\
  ed.\BibitemShut {Stop}%
\bibitem [{\citenamefont {Worth}\ \emph {et~al.}(2016)\citenamefont {Worth},
  \citenamefont {Beck}, \citenamefont {J{\"a}ckle}, \citenamefont {Vendrell},\
  and\ \citenamefont {Meyer}}]{mctdh_package_ml}%
  \BibitemOpen
  \bibfield  {author} {\bibinfo {author} {\bibfnamefont {G.~A.}\ \bibnamefont
  {Worth}}, \bibinfo {author} {\bibfnamefont {M.~H.}\ \bibnamefont {Beck}},
  \bibinfo {author} {\bibfnamefont {A.}~\bibnamefont {J{\"a}ckle}}, \bibinfo
  {author} {\bibfnamefont {O.}~\bibnamefont {Vendrell}}, \ and\ \bibinfo
  {author} {\bibfnamefont {H.-D.}\ \bibnamefont {Meyer}},\ }\href@noop {}
  {}\bibinfo {howpublished} {The {MCTDH} {P}ackage, {V}ersion 8.5.6.1. {S}ee
  http://mctdh.uni-hd.de} (\bibinfo {year} {2016})\BibitemShut {NoStop}%
\end{thebibliography}
\end{document}